\documentstyle[12pt]{article}

\renewcommand{\theequation}{\arabic{section}.\arabic{equation}}

\newcommand{\be}{\begin{equation}}
\newcommand{\ee}{\end{equation}}
\newcommand{\bea}{\begin{eqnarray}}
\newcommand{\eea}{\end{eqnarray}}

 at 14.4truept
 at 14.4truept

\def\pa{\partial}
\def\ker{{{\rm Ker}({\rm ad\,} \Lambda)}}

\def\H{{\cal H}}
\def\R{{\bf R}}

\def\k{{\bar k}}
\def\tr{{\rm tr\,}}
\def\M{{\cal M}}

\def\W{{\cal W}}
\def\G{{\cal G}}
\def\L{{\cal L}}
\def\M{{\cal M}}

\def\N{{\cal N}}

\def\1{{ 1}}
\def\cinf{{C^\infty}}
\begin{document}

\thispagestyle{empty}

\setcounter{page}{0}
\renewcommand{\thefootnote}{\fnsymbol{footnote}}
\fnsymbol{footnote}
\phantom{NOTHING}
\rightline{ENSLAPP-L-493/94\break}
\rightline{hep-th/9410203\break}
\rightline{revised version}
\rightline{SWAT-95-77}

\vspace{.4in}

\begin{center} \Large \bf
Regular Conjugacy Classes in the Weyl Group\\

and Integrable Hierarchies
\end{center}

\vspace{.1in}

\begin{center}
F.~Delduc
and L.~Feh\'er${}$\protect\footnote{
Present address: Department of Physics, University of Swansea,
Singleton Park,
Swansea SA2 8PP, UK.
\phantom{On.}
On leave from Theoretical Physics Department  of Szeged University, H-6720
Szeged, Hungary.}\\

\vspace{0.2in}

{\em Laboratoire de Physique Th\'eorique\footnote{\rm ENSLAPP, URA 14-36 du
CNRS, associ\'e \`a l'E.N.S. de Lyon et au L.A.P.P.}\\
ENS Lyon,  46 all\'ee d'Italie\\
F-69364 Lyon Cedex 07, France}
\end{center}

\vspace{.2in}

{\parindent=25pt
\narrower\smallskip\noindent

\small
\noindent
{\bf Abstract.}\quad
Generalized KdV hierarchies  associated by Drinfeld-Sokolov reduction to grade
one regular  semisimple  elements from non-equivalent  Heisenberg subalgebras
of a loop algebra  $\G\otimes{\bf C}[\lambda,\lambda^{-1}]$ are studied.
The graded Heisenberg subalgebras  containing such elements  are labelled by
the regular conjugacy  classes in the Weyl group ${\bf W}(\G)$ of the simple
Lie algebra $\G$. A representative $w\in {\bf W}(\G)$ of a regular conjugacy
class can be lifted to an inner automorphism of $\G$  given by
$\hat w=\exp\left(2i\pi {\rm ad I_0}/m\right)$,
where $I_0$ is the defining vector of an $sl_2$ subalgebra of $\G$.
The grading is then defined  by the operator
$d_{m,I_0}=m\lambda {d\over d\lambda} + {\rm ad} I_0$
and any grade  one regular element $\Lambda$ from the Heisenberg
subalgebra associated to $[w]$ takes the form $\Lambda = (C_+ +\lambda C_-)$,
where $[I_0, C_-]=-(m-1) C_-$ and $C_+$ is included in an $sl_2$ subalgebra
containing $I_0$. The largest eigenvalue of ${\rm ad}I_0$ is $(m-1)$ except
for some cases in $F_4$, $E_{6,7,8}$.
We explain how these Lie algebraic results follow from known
results and  apply them to construct integrable systems.
If the largest ${\rm ad} I_0$ eigenvalue is $(m-1)$, then using
any grade one regular element from the  Heisenberg subalgebra
associated to $[w]$ we can construct a KdV system possessing the standard
$\W$-algebra defined by $I_0$ as its second Poisson bracket algebra.
For $\G$ a classical Lie algebra,
we derive  pseudo-differential  Lax operators for those
non-principal KdV systems that can be  obtained as
discrete reductions
of KdV systems related to $gl_n$.
Non-abelian Toda systems are also considered.
}

\normalsize

\newpage
\renewcommand{\thefootnote}{\arabic{footnote}}
\setcounter{footnote}{0}

\vspace*{4.0cm}
\begin{center}{\Large\bf Contents}\end{center}
\vspace*{1.0cm}
\contentsline {subsection}{\numberline {1}Introduction}{2}
\vspace*{0.4cm}
\contentsline {subsection}{\numberline {2}Heisenberg subalgebras and
the Weyl group}{7}
\vspace*{0.4cm}
\contentsline {subsection}{\numberline {3}Regular conjugacy classes in
the Weyl group}{9}
\vspace*{0.1cm}
\contentsline {subsection}{\numberline {3.1}Regular conjugacy classes in
${\bf W}(A_{n-1})$}{9}
\contentsline {subsection}{\numberline {3.2}Regular conjugacy classes in
${\bf W}(D_n)$}{10}
\contentsline {subsection}{\numberline {3.3}Regular conjugacy classes in
${\bf W}(B_n)\simeq {\bf W}(C_n)$}{12}
\vspace*{0.4cm}
\contentsline {subsection}{\numberline {4}Heisenberg subalgebras with
graded regular elements and $sl_2$ embeddings}{13}
\vspace*{0.1cm}
\contentsline {subsection}{\numberline {4.1}A practical algorithm to
construct Heisenberg subalgebras}{13}
\contentsline {subsection}{\numberline {4.2}A connection to $sl_2$
embeddings}{16}
\vspace*{0.4cm}
\contentsline {subsection}{\numberline {5}Applications to KdV
type systems}{21}
\vspace*{0.1cm}
\contentsline {subsection}{\numberline {5.1}KdV systems associated to
 grade one regular elements}{21}
\contentsline {subsection}{\numberline {5.2}Examples: Lax operators of
Gelfand-Dickey type}{25}
\vspace*{0.4cm}
\contentsline {subsection}{\numberline {6}Some remarks on non-abelian
 Toda systems}{34}
\vspace*{0.4cm}
\contentsline {subsection}{\numberline {7}Conclusion}{37}
\vspace*{0.4cm}
\contentsline {subsection}{\numberline {A}Canonical $sl_2$ for any
regular primitive conjugacy class}{39}
\vspace*{0.4cm}
\contentsline {subsection}{References}{42}

\newpage

\section{Introduction}

The purpose of this paper is to contribute to
the classification  of generalized KdV systems that may be obtained from
the Drinfeld-Sokolov approach to integrable hierarchies.
One of the main achievements presented in the seminal paper \cite{DS}
by Drinfeld  and Sokolov was  the interpretation in terms of affine
Lie algebras of the $n$-KdV hierarchies defined by Gelfand and Dickey
in \cite{GD,Di}
and Adler in \cite{Ad} in terms of the calculus of pseudo-differential
 operators.
The phase space consisting of scalar Lax operators
\begin{equation}
L=\pa^n + u_1 \pa^{n-1}+\cdots +u_{n-1}\pa + u_n,
\qquad
u_i\in\cinf(S^1,{\bf C}),
\label{1.1}\end{equation}
was interpreted as the reduced phase space following a Hamiltonian
symmetry reduction applied to the dual of an affine Lie algebra.
This explained the origin of the quadratic
Adler-Gelfand-Dickey Poisson bracket as a reduced Lie-Poisson bracket
and also explained the commuting Hamiltonians generated by residues
of fractional powers of $L$
as being reductions of those obtained by applying
the Adler-Kostant-Symes
scheme  to the affine Lie algebra (see also \cite{RSTS}).
The properties of the matrix
\begin{equation}
\Lambda_n
=\left[\matrix{
0&1&0&\cdots&0\cr
\vdots&0&1&\ddots&\vdots\cr
\vdots&{}&\ddots&\ddots&0\cr
0&{}&{}&\ddots&1\cr
\lambda &0&\cdots&\cdots&0\cr}\right]
\label{1.2}\end{equation}
played a crucial role  in the construction.
The centralizer of $\Lambda_n$ in the loop algebra
$\ell(gl_n):= gl_n\otimes {\bf C}[\lambda,\lambda^{-1}]$ is a graded
maximal
abelian subalgebra, which becomes the principal Heisenberg subalgebra
upon central extension \cite{kac}.
The commuting flows were constructed out of this abelian subalgebra
making essential use of the principal grading and the
regularity of the element $\Lambda_n$ that has grade one.
The other main achievement of Drinfeld and Sokolov was
the derivation of new KdV type  hierarchies by generalizing the
construction to an arbitrary affine Lie algebra using the
respective principal Heisenberg subalgebra and its grade one
regular element.
Like the KdV type systems of \cite{DS},
the affine Toda systems are also based on
the principal Heisenberg
subalgebra, with the grading and the regular element of grade one
playing an important role.

The generalized KdV systems that will be studied in this paper
will be associated to regular elements of grade one from certain
non-principal Heisenberg subalgebras of
$\ell(\G):=\G\otimes [\lambda,\lambda^{-1}]$ for $\G$ a
simple Lie algebra using the Hamiltonian reduction technique of
\cite{DS}.
Related non-abelian affine Toda systems will be also presented.

Generalizations of
the Drinfeld-Sokolov construction of integrable hierarchies
have  already been considered in the literature.
Soon after \cite{DS},
Wilson  \cite{Wi} suggested associating
systems of modified KdV  and Toda type to any grade one semisimple
element of any affine Lie algebra,
with respect to a grading
defined by an automorphism of finite order of the corresponding finite
dimensional simple Lie algebra.
In the context of Toda field theories,
similar proposals can be found in \cite{LS,Oli,under}.
Concerning the important,  apparently still open,
problem of  classifying the gradings that admit a grade one
semisimple element,
some progress was made in \cite{McI,under}.
The construction of systems of modified KdV type
can be done without any reference to a gauge freedom,
while the presence of a non-trivial gauge freedom
is a crucial  ingredient in the construction of the KdV
type systems in \cite{DS}.
In  the unpublished work \cite{McI},
the reduction procedure of \cite{DS} was  generalized in order to obtain
generalized Miura maps for associating KdV
type systems to those of modified KdV type.
It was also realized in \cite{McI} that the semisimple
element and the gradings involved in the generalized
Drinfeld-Sokolov reduction must satisfy a certain non-degeneracy condition,
which is required for the existence of the global, polynomial gauges
that define the KdV fields as in \cite{DS}.
More recently,  the ideas of \cite{Wi}
were resurrected and made concrete by de Groot et al
\cite{Prin1,Prin2,Prin3,SdeC}
 taking advantage of the theory of
non-equivalent graded Heisenberg subalgebras in the affine
Lie algebras developed by Kac and Peterson  \cite{KP}.
In \cite{Prin1} it was suggested  to use any graded
element  $\Lambda$ with non-zero grade from
any Heisenberg subalgebra of an affine Lie
algebra in a generalized Drinfeld-Sokolov reduction procedure.
Such an element  $\Lambda$
is automatically semisimple and
in \cite{Prin1} two types of systems, called type I and type II,
were distinguished according to whether $\Lambda$ is regular
or non-regular. The notion of regularity is defined  below.
In the type I cases it is possible
to verify the existence of  the polynomial gauges
(``DS gauges'') required for the construction of KdV type systems.
This in general is not so in the type II cases and has
to be imposed as an extra condition for obtaining  KdV type systems.

In fact the approach used in \cite{Prin1} is
almost the same as the one  in \cite{McI}.
In the setup of \cite{Prin1} the semisimple element
$\Lambda$ can have any non-zero grade,
but in the most interesting cases when $\Lambda$ has grade one the
two methods almost always coincide.
Indeed in the case of the classical simple Lie algebras
we are  aware of no exceptions.
An advantage of the  approach used in \cite{Prin1}
is that it incorporates  a universal definition of the gauge group
which  is applicable to any graded semisimple element
$\Lambda$ and  implies the existence of polynomial gauge fixings
if $\Lambda$ is regular.

According to the above, one can associate
generalized KdV systems to certain graded semisimple elements of the affine
Lie algebras that include the regular elements of minimal
non-zero (say positive) grade taken from the non-equivalent graded
Heisenberg subalgebras.
It appears a reasonable strategy to first explore  the systems
that may be associated to the non-equivalent regular semisimple
elements of minimal grade.
Progress in  this direction was reported in \cite{FHM,FM}, where the
case  of  the affine Lie algebra $\ell(gl_n)$ was considered.
In this case the graded Heisenberg subalgebras
are classified by the partitions on $n$ \cite{KP,ten}  and
it was verified in \cite{FHM} that
only the partitions of $n$ into sums of equal numbers,
$n=sp$, and into sums of equal numbers plus one,
$n=sp+1$, admit a graded regular element.
A generalized Drinfeld-Sokolov  reduction
based on a grade one regular element from the Heisenberg subalgebra
associated to the partition $n=sp$ was analyzed in \cite{FHM}
and was found to lead to the matrix version of the
Gelfand-Dickey hierarchy given by Lax  operators of the form
\begin{equation}
L= Q \pa^p + u_1 \pa^{p-1}
+ \cdots
+u_{p-1} \pa+  u_p,
\qquad
u_i\in \cinf(S^1,gl_s),
\label{1.3}\end{equation}
where $Q$ is a diagonal constant matrix with distinct, non-zero entries.
 In the case $n=sp+1$
the analogous  Drinfeld-Sokolov  reduction (see \cite{FM})  yields  a
hierarchy associated to
a more exotic
 looking $s\times s$ matrix
Lax operator:
\begin{equation}
L= Q \pa^p + u_1 \pa^{p-1}
+ \cdots
+u_{p-1} \pa +  u_p - y_+ (\pa + w )^{-1} y^t_-\,,
\label{1.4}\end{equation}
where the fields $u_i$ vary like in (\ref{1.3}),
 $y_\pm\in\cinf( S^1,{\bf C}^s)$ and $w\in\cinf( S^1,{\bf C})$.
For the history of this model and for related  recent
developments on KdV type hierarchies,
the reader may consult refs.~\cite{Cheng,OS,Deck,Dic,Bon,Ara},
in all of which different methods to those in \cite{FHM,FM}
were used.

In  none of the above mentioned papers had it been  realized
that a  classification of the graded regular semisimple elements
of the affine Lie algebras can be extracted from known results.
We now explain this in the non-twisted case.
Let $\G$ be a complex simple Lie algebra.
Disregarding the central extension,
recall from \cite{KP} that
the graded Heisenberg subalgebras of the non-twisted
 loop algebra $\ell(\G)$ are classified by
the conjugacy classes (see \cite{cart})
in the Weyl group ${\bf W}(\G)$ of $\G$.
It is also clear from the construction in \cite{KP} that
the  graded regular elements in a  Heisenberg subalgebra,
$\tilde {\cal H}_{\hat w}\subset \ell(\G)$
associated to the conjugacy class $[w]\subset {\bf W}(\G)$,
correspond to the regular eigenvectors
of the  Weyl transformation $w\in [w]$ acting
on the Cartan subalgebra $\H\subset \G$.
In \cite{Sp} the conjugacy classes in the Weyl group whose representatives
admit a regular eigenvector
(an eigenvector whose centralizer in $\G$ is $\H$)
are themselves called regular.
The  regular conjugacy classes in the Weyl groups
were then  all classified by Springer \cite{Sp}.
This yields a  classification of the  graded regular semisimple
elements of $\ell(\G)$, since every such element is
contained in a  graded Heisenberg subalgebra.
Although this classification is not yet complete since there
are ambiguities in choosing the grading
of $\ell(\G)$ associated to  a  conjugacy class $[w]\subset {\bf W}(\G)$,
because the construction involves lifting a representative $w\in [w]$
to a finite order inner automorphism
$\hat w=\exp\left(2i\pi {\rm ad} X\right)$ of $\G$,
we shall see that there exists a  natural choice for every regular
conjugacy class.

In this paper the above classification of the graded regular
semisimple elements of the loop algebras $\ell(\G)$ will be
developed and  applications will be considered concentrating on the
classical simple Lie algebras.
In addition to the theory of integrable systems,
our work is also motivated by the relations
between  integrable hierarchies
and various other subjects of two dimensional theoretical physics,
${\cal W}$-algebras  and 2d gravity models
being prime examples (e.g.~\cite{Zam,Doug,bais,rep,BS,Mor}).
An important question for us is
to clarify the relationship between generalized KdV hierarchies and
$\W$-algebras,
which is well-known in the original Drinfeld-Sokolov case.
We will be able to associate a KdV type hierarchy to every  grade one
regular element from a graded Heisenberg subalgebra
of $\ell(\G)$ in such a way that the second
Poisson bracket of the hierarchy gives a classical $\W$-algebra
associated to a corresponding  $sl_2$ subalgebra of $\G$.
The set of $\W$-algebras arising in this way is a small
subset of the standard $\W$-algebras associated to arbitrary $sl_2$
embeddings \cite{bais,rep}.
Our result on the $\W$-algebra structures corresponding
to  the KdV systems  is consistent with the results in
\cite{SdeC}, where a $\W$-subalgebra
was exhibited in the second Poisson bracket algebra for
a certain  class  of generalized KdV hierarchies.
By the method of \cite{Prin1,Prin2},
these hierarchies are
associated to a graded semisimple element  $\Lambda$
subject to a certain non-degeneracy condition,
which is satisfied in all the  cases that we
shall consider.

Before describing the content of the paper in more detail,
it is worthwhile to recapitulate the essence of the use of a graded
regular semisimple element of non-zero grade to
integrable systems in technical
terms.
An element $\Lambda$ of a non-twisted loop algebra
$\ell({\cal G})$,
 where ${\cal G}$ is a simple Lie algebra or $gl_n$,
is called {\it semisimple}  if it defines a direct sum decomposition
\begin{equation}
\ell({\cal G})={\rm Ker}({\rm ad\,} \Lambda) + {\rm Im}({\rm ad\,} \Lambda).
\label{1.5}\end{equation}
By definition, a  semisimple element $\Lambda$ is
{\em regular} if
$\mbox{Ker}({\rm ad\,} \Lambda)\subset \ell(\G)$  is an
{\it abelian} subalgebra.
The ${\bf Z}$-grading in which $\Lambda$  is
supposed to be homogeneous with non-zero grade  is defined by the
eigenspaces of a  linear operator
$d_{N,Y}: \ell({\cal G}) \rightarrow \ell({\cal G})$,
\begin{equation}
d_{N,Y} = N \lambda {d\over d\lambda} + {\rm ad\,} Y ,
\label{1.6}\end{equation}
where $N$ is a non-zero integer and $Y\in {\cal G}$ is diagonalizable
with integer eigenvalues in the adjoint representation.
If one has such an element, then
 ${\rm Ker}({\rm ad\,} \Lambda)$ is a {\it graded, maximal} abelian
subalgebra.
Note also that  ${\rm ad\,} Y$ defines a  grading
${\cal G} = \oplus_i {\cal G}_i$ of ${\cal G}$. The most
important graded regular semisimple elements are of small grade taking
the form
\begin{equation}
\Lambda = C_+ + \lambda C_-
\qquad
\hbox{with some}\qquad
C_\pm \in {\cal G}.
\label{1.7}\end{equation}
The integrable hierarchies of our interest are given by Hamiltonian
flows on a phase space consisting of
first order differential
operators ${\cal L}$ of the type
\begin{equation}
{\cal L} = \pa + j + \Lambda
\qquad\hbox{with}\quad
 j: S^1 \rightarrow \sum_{i<k} \ell(\G)_i,
\label{1.8}\end{equation}
where $\ell(\G)_i\subset \ell(\G)$ is the grade $i$ eigensubspace of
$d_{N,Y}$ and $k>0$ is the grade of $\Lambda$.
In addition to being restricted to grades  strictly smaller
than the grade  of the leading  term $\Lambda$,
the field $j$ in (\ref{1.8}) is usually also subject to
further constraints (e.g.~it often varies in $\G\subset \ell(\G)$ only)
and to a gauge freedom specific to the system.
Since the field $j$ is periodic (being a function on the space $S^1$),
one can consider the monodromy matrix  of ${\cal L}$.
The point is that under the above assumptions  one may obtain
commuting {\em local}  Hamiltonians from the monodromy invariants
determined by  the  ``abelianization'' of ${\cal L}$
\cite{DS,Wi,Oli,McI,Prin1}.
This abelianization is essentially a perturbative diagonalization
which is achieved by transforming ${\cal L}$  (\ref{1.8}) according to
\begin{equation}
\left({\pa + j +\Lambda }\right)
\mapsto e^{{\rm ad\,} F } \left({\pa + j +\Lambda }\right)
 := (\pa + h + \Lambda),
\label{1.9a}\end{equation}
where $F$ and $h$ are infinite series required to take their values
in appropriate graded subspaces in the decomposition (\ref{1.5}):
\begin{equation}
F: S^1 \rightarrow \left({\rm Im}({\rm ad\,} \Lambda)\right)_{<0} ,
\qquad
h: S^1\rightarrow \left({\rm Ker}({\rm ad\,} \Lambda)\right)_{<k}.
\label{1.9b}\end{equation}
In fact, the above assumptions ensure that (\ref{1.9a}), (\ref{1.9b})
can be solved recursively, grade by grade, for both $F(j)$ and $h(j)$
and the solution is given by
unique {\it differential polynomials} in the components of $j$.
The local monodromy invariants are  the integrals over $S^1$
of the graded components of the resulting $h(j)$.
In an appropriate  hamiltonian setting, these provide the
Hamiltonians that generate a
hierarchy  of commuting  evolution equations.

\smallskip

The rest of this paper is organized as follows.
Sections 2, 3 and 4 are devoted to presenting some Lie algebraic results
relevant for the classification of generalized KdV systems.
In Section 2 it is explained that
the   classification  of the graded regular semisimple elements of
a loop algebra $\ell(\G)$ can be reduced to the classification of
the regular eigenvectors of representatives of
the conjugacy classes in the Weyl group ${\bf W}(\G)$ of $\G$
thanks to results in \cite{KP}.
The solution of this classification problem which is
due to Springer \cite{Sp},
is summarized in Tables 1, 2 and 3  in Section 3 for $\G$ a classical
simple  Lie algebra.

In Section 4 we describe a connection between the regular conjugacy
classes in ${\bf W}(\G)$, with associated grade one regular
semisimple elements in $\ell(\G)$, and certain $sl_2$ subalgebras
in the classical Lie algebra $\G$.
For every regular conjugacy class $[w]\subset {\bf W}(\G)$ of order $m$,
we shall exhibit a lift $\hat w$ of a representative $w\in [w]$
having the form
\begin{equation}
\hat w=\exp\left( 2i\pi {\rm ad} I_0/m\right),
\label{1.11}
\end{equation}
where $I_0$ is the  defining vector \cite{dyn}  of
an $sl_2$ subalgebra of $\G$ and the largest eigenvalue
of ${\rm ad} I_0$ is $(m-1)$.
The order of the inner automorphism $\hat w$ of $\G$ is
$\nu m$, where $\nu$ is $1$ or $2$ depending on whether
${\rm ad} I_0$ has only integral or also half-integral eingenvalues.
Actually $\nu=1$ in  almost all cases.
Using this $\hat w$ in the Kac-Peterson construction
of the graded Heisenberg subalgebra,
$\tilde {\cal H}_{\hat w}\subset \ell(\G)$
associated to $[w]\subset {\bf W}(\G)$,
induces the  ${\bf Z}/\nu$ grading on $\ell(\G)$
defined by the operator
$d_{m,I_0}=m\lambda {d\over d\lambda} +{\rm ad} I_0$.
This is the natural grading of $\ell(\G)$ which we associate to $[w]$.
We then show that every graded regular element
$\Lambda \in \tilde {\cal H}_{\hat w}$ of minimal positive
grade, in fact $d_{m,I_0}$ grade one, has the form (1.7),
{\em where $C_+$  can be included in an $sl_2$ subalgebra
containing also $I_0$.}
That is there exists $I_-\in \G$ for which $[I_0, I_\pm ]=\pm I_\pm$,
$[I_+, I_-]=2I_0$ holds with $I_+:=C_+$ contained in
 $\Lambda=(C_+ +\lambda C_-)$.

The above connection between regular conjugacy classes in ${\bf W}(\G)$
and $sl_2$ subalgebras in $\G$ generalizes and in many cases is
implied by  the classical result of Kostant \cite{Kost} on the
connection between the Coxeter class in ${\bf W}(\G)$ and the
principal $sl_2$ subalgebra in  $\G$.
In the main text we shall take $\G$ to be a classical Lie algebra,
but in Appendix A we discuss the connection between regular conjugacy
classes in ${\bf W}(\G)$ and $sl_2$ embeddings in $\G$ for an arbitrary
simple Lie algebra too.
In the algebras  $F_4$ and $E_{6,7,8}$ we find that $(m-1)$ in (1.11) is
smaller than the largest ${\rm ad} I_0$ eigenvalue in some cases,
but equality holds for every {\em regular primitive} conjugacy class.
As will be clear from our references, we
do not claim credit for original group theoretic results.
However, by inspecting and systematizing a number of isolated results,
we will be  able to formulate and verify interesting general  statements,
which are worth knowing  but to our knowledge are
not available in the literature.

We turn to the application of the above results  to
the construction of KdV
type integrable hierarchies in Section 5.
In Subsection 5.1 we associate a KdV type system
to every grade one  regular semisimple element
$\Lambda \in \tilde {\cal H}_{\hat w}$.
This hierarchy will be obtained by a direct generalization of the
standard Drinfeld-Sokolov reduction.
We assume that  the largest
eigenvalue of ${\rm ad} I_0$  equals $(m-1)$ in (1.11),
which is always  satisfied
if $\G$ is a classical simple Lie algebra or $G_2$.
The  second Poisson bracket algebra of the resulting
generalized KdV hierarchy is then
the $\W$-algebra \cite{bais,rep}
belonging to the $sl_2$ embedding defined by $I_0$.
In Subsection 5.2 we derive
Gelfand-Dickey type Lax operators for a subset of the
generalized KdV systems.
These systems  correspond  to conjugacy
classes in the Weyl group of a classical Lie
algebra  given by the product of Coxeter elements in a regular
subalgebra composed of $A$ and $C$ type simple factors.
They   turn out to be ``discrete reductions'' of
generalized KdV systems related  to $gl_n$  given by
Lax operators of the form in (\ref{1.3}) and (\ref{1.4}).
In Section 6 we briefly comment on  non-abelian affine Toda systems
and present the detailed form of the
non-abelian affine Toda equation corresponding  to
the regular, primitive (semi-Coxeter) conjugacy class
$(\bar p,\bar p)\subset  {\bf W}(D_{2p})$.

Finally,  we give our conclusions and comment on some
open problems in Section 7.

\newpage

\section{Heisenberg subalgebras and the Weyl group}
\setcounter{equation}{0}

Let $\G$ be a complex simple Lie algebra.
Consider  the Lie algebra $\ell(\G)$ of Laurent polynomials,
$\ell(\G):=\G\otimes {\bf C}[\lambda,\lambda^{-1}]$,
in the spectral parameter $\lambda$.
For any graded regular semisimple element
$\Lambda \in \ell({\G})$,
$\ker\subset \ell(\G)$ is a graded maximal abelian subalgebra, which
becomes a Heisenberg subalgebra upon centrally   extending $\ell(\G)$.
In order to find the  graded regular
semisimple  elements of $\ell(\G)$,
it is therefore enough to inspect the
maximal abelian subalgebras of $\ell(\G)$ that underlie the
graded Heisenberg subalgebras of the central extension
$\widehat\G$ of $\ell(\G)$, and select those which contain
graded regular elements.
With respect to the adjoint action of an
appropriate group associated to $\ell(\G)$,
the non-equivalent
graded Heisenberg subalgebras of $\widehat \G$
are {\em classified} by the conjugacy classes in the Weyl group
of $\G$ \cite{KP}.
See also \cite{segal,fons} for the precise statement.
Next we recall the main points of the construction
on which this classification is based.
Note that, by disregarding the central extension,
a maximal abelian subalgebra of $\ell(\G)$ will be  often referred to
as a Heisenberg subalgebra throughout the text.

Suppose that $\H\subset \G$ is a Cartan subalgebra and $\tau$ is a
{\it finite order, inner}  automorphism of $\G$ that normalizes $\H$.
Consider the  following models of $\ell(\G)$ and its twisted
realization $\ell(\G,\tau)$:
\begin{eqnarray}
&\ell(\G)=\{ F\,\vert\, F: \R\rightarrow \G,
\quad F(\theta+2\pi)=F(\theta)\,\},\nonumber\\
& \ell(\G,\tau)=\{ f\,\vert\, f: \R\rightarrow \G,
\quad
f(\theta+2\pi)=\tau\left(f(\theta)\right)\,\}.
\label{2.1}\end{eqnarray}
Since $\tau$ is inner, $\ell(\G)$ and $\ell(\G,\tau)$ are
isomorphic \cite{kac,hel}.
To see this one writes $\tau$ as
\begin{equation}
\tau = e^{2 i \pi  {{\rm ad } X} },
\quad
X={Y/N},
\label{2.2}\end{equation}
where $N$ is the order of $\tau$, $\tau^N = {\rm id}$, and $Y\in \G$
is diagonalizable. The choice of $Y$ is not unique.
The isomorphism  $\eta: \ell(\G, \tau)\rightarrow \ell(\G)$
is given by ``untwisting'' as follows:
\begin{equation}
\eta: f\mapsto F\,,
\qquad
F(\theta):=e^{- i \theta  {{\rm ad\,} X}}\left(f(\theta)\right).
\label{2.3}\end{equation}
The ``twisted homogeneous Heisenberg subalgebra'' $\ell(\H,\tau)$,
\begin{equation}
\ell(\H,\tau)=\{ f\,\vert\, f: \R\rightarrow \H,
\quad
f(\theta+2\pi)=\tau\left(f(\theta)\right)\,\},
\label{2.4}\end{equation}
is a  maximal abelian subalgebra of $\ell(\G,\tau)$.
The image ${\tilde \H}_\tau:=\eta[\ell(\H,\tau)]$
of the  twisted homogeneous Heisenberg subalgebra is a maximal
abelian subalgebra of $\ell(\G)$.
The natural grading on $\ell(\G,\tau)$ is the
homogeneous grading defined by the eigensubspaces of
$d: \ell(\G,\tau)\rightarrow \ell(\G,\tau)$,
\begin{equation}
d:= -i N {d\over d\theta}.
\label{2.5}\end{equation}
The isomorphism $\eta$ induces a corresponding grading operator
$d_{N,Y}: \ell(\G)\rightarrow \ell(\G)$,
\begin{equation}
d_{N,Y} := \eta \circ d \circ \eta^{-1}=N\lambda {d\over d\lambda}
+{\rm ad\,} Y\,,
\label{2.6}\end{equation}
where we used  the definition $\lambda:=e^{i\theta}$.
The  maximal abelian subalgebras
$\ell(\H,\tau)\subset \ell(\G,\tau)$ and
${\tilde \H}_\tau\subset \ell(\G)$ are of course graded.

\newpage 

Recall (e.g.~\cite{hel}) that Weyl group
${\bf W}(\G)$ of $\G$ may be identified as the group
of inner automorphisms of $\G$ that normalize $\H$ modulo
the inner automorphisms centralizing $\H$.
It is also well-known
that any $w\in {\bf W}(\G)$ may be, in general non-uniquely,  lifted to a
{\em finite order} inner automorphism $\hat w$ of $\G$
which reduces  to $w$ on  $\H$, $\hat w\vert_{\cal H}=w$.
It follows that one can  associate a graded maximal abelian
subalgebra,  ${\tilde \H}_{\hat w}\subset \ell(\G)$,
to any element $w\in {\bf W}(\G)$.
To construct ${\tilde \H}_{\hat w}\subset \ell(\G)$,
one first lifts $w\in [w]$  and
then performs the above construction using $\hat w$
in place of $\tau$ in (2.1)--(2.6).
Despite the  ambiguities involved, it can be shown
\cite{KP,segal,fons} that conjugate elements of ${\bf W}(\G)$
give rise to equivalent graded Heisenberg subalgebras
and the non-equivalent ones are classified by the conjugacy
classes in ${\bf W}(\G)$.

We now need to construct a graded basis of $\ell(\G,\hat w)$.
This is  done as follows.
The eigenvalues of $\hat w$ on $\G$ are of the form $\omega^\k$ with
\begin{equation}
\omega:=\exp\left(2i \pi / N\right)
\qquad \hbox{and}\qquad
\k\in \{0, 1,\ldots, (N-1)\},
\label{2.7}\end{equation}
where $N$ is the order of $\hat w$.
A basis of $\G$ consisting of eigenvectors of $\hat w$  may
be given in the form $\{ H_{\k, q_\k}\}\cup \{ R_{\k,r_\k}\}$
with
\begin{equation}
w(H_{\k,q_\k})=\omega^\k H_{\k,q_\k},
\ \
H_{\k,q_\k}\in {\cal H}
\qquad\hbox{and}\qquad
\hat w(R_{\k, r_\k})=\omega^\k R_{\k,r_\k}\,,
\ \
R_{\k,r_\k} \in {\cal H}^\perp,
\label{2.8}\end{equation}
that is by separately diagonalizing $\hat w$ on
the Cartan subalgebra $\H$ (where it reduces to $w$)
and on
its complementary space
$\H^\perp \subset \G$ spanned by the root vectors.
The index  $q_\k$, similarly $r_\k$,  counts the multiplicity
of the corresponding  eigenvalue, which can be also zero of course.
The desired graded basis of $\ell(\G,\hat w)$ consists of
the elements
\begin{equation}
z^k H_{\k,q_\k}
\quad\hbox{and}\quad
z^k R_{\k,r_\k}
\quad\hbox{where}\quad
z:=\exp\left({ i \theta/ N}\right) ,
\quad
k=\k\ {\rm mod}\  N.
\label{2.9}\end{equation}
By definition,
a graded element $z^k H_{\k,q_\k} \in \ell(\H,\hat w)
\subset \ell(\G,\hat w)$ of grade
$k$ is  {\it regular}  if
\begin{equation}
\ell(\G,\hat w)\supset {\rm Ker}\left({\rm ad\,} z^k H_{\k,q_\k} \right)
= \ell(\H,\hat w).
\label{2.10}\end{equation}
It is  easy to see that (\ref{2.10}) is equivalent to
\begin{equation}
\G \supset
{\rm Ker}\left( {\rm ad\,} H_{\k,q_\k}\right) ={\cal H}.
\label{2.11}\end{equation}
Equations (2.10) and (2.11) refer respectively to infinite
and finite dimensional Lie algebras.
Using  standard terminology  in the finite dimensional case,
$H\in {\cal H}$ is by definition
  {\em regular} if its centralizer in $\G$ is $\H$.
Hence the equivalence of (2.10) and (2.11)  means that
{\em $z^k H_{\k,q_\k}\in \ell(\H, \hat w)$ is a regular semisimple element
of $\ell(\G,\hat w)$
if and only if $H_{\k, q_\k}\in \H$ is a regular semisimple element
of $\G$}.
In principle,
this simple statement should make it possible to find all
graded regular semisimple elements of $\ell(\G)$.

In order to find the  graded regular semisimple
elements of $\ell(\G)$,   one  needs  to select the conjugacy classes
$[w]\subset {\bf W}(\G)$ for which
the graded maximal abelian subalgebra
${\tilde \H}_{\hat w}\subset \ell(\G)$
contains a graded regular element.
By the isomorphism between $\ell(\G, \hat w)$ and $\ell(\G)$
that brings $\ell(\H,\hat w)$ into $\tilde \H_{\hat w}$
and the  statement above, this problem
is equivalent  to selecting the conjugacy classes in ${\bf W}(\G)$
whose representatives  admit a regular eigenvector.
A conjugacy class with this property is called
a {\it regular conjugacy class} in \cite{Sp}, where
all such conjugacy classes have been listed.

\medskip

\noindent{\em Remark.}
It is apparent  from
the above construction of the Heisenberg subalgebra
$\tilde {\cal H}_{\hat w}\subset  \ell(\G)$
associated  to $[w]\subset {\bf W}(\G)$ that the corresponding
grading of $\ell(\G)$
depends on the choice of the finite order inner automorphism
$\hat w$ used for defining  the lift   of a representative $w\in [w]$.
As the grading plays a crucial role in the Drinfeld-Sokolov construction,
a  clarification of this ambiguity,
in terms of the classification  of finite order automorphisms due to
Kac  \cite{kac,hel},  would be desirable.
This problem will not be addressed in the present paper.
Rather, in  Section 4 and in Appendix A,
a  distinguished  lift having the  nice properties in (\ref{1.11})
will be exhibited for every regular conjugacy class in the Weyl group.

\section{Regular conjugacy classes in the Weyl group}
\setcounter{equation}{0}

The conjugacy classes in the Weyl group
are described in \cite{cart}
for all simple Lie algebras, and
the regular  conjugacy  classes (which admit  a regular eigenvector)
are described  in \cite{Sp}.
In this section we recall the relevant results of \cite{Sp}
in the form of tables for the classical simple Lie algebras,
which will be used in our applications later.
In these tables we shall also present the explicit form of the
regular eigenvectors for convenient representatives of the regular
conjugacy classes.
The eigenvectors are not given in \cite{Sp}, but can be
easily computed.
As a matter of fact the classification of the regular conjugacy classes
can be  also derived straightforwardly by explicitly
diagonalizing a  representative for each conjugacy class
and  inspecting the eigenvectors.
In our study originally we  used this ``brute
force'' approach, but after learning the elegant work of Springer
\cite{Sp} this explicit inspection became
superfluous and will not  be presented apart from some remarks.
By means of the natural scalar product,
the Cartan subalgebra
$\cal H\subset \G$ will be always identified with the space of roots
${\cal H}^*$  in this section.

\subsection{ Regular conjugacy classes in ${\bf W}(A_{n-1})$}

The Cartan subalgebra of $A_{n-1}$  may be identified with
the subspace of the vector space spanned by $n$ orthonormal vectors
$\epsilon_l$,
$l=1,\ldots,n$ which is orthogonal to the vector $\sum_{l=1}^n \epsilon_l$.
The roots of $A_{n-1}$ are the vectors $\epsilon_l-\epsilon_{l'}$,
$l\neq l'$.
An element
\begin{equation}
H = \sum_{l=1}^n h_l\epsilon_l,\quad \sum_{l=1}^n h_l=0,
\label{3.1}\end{equation}
of the Cartan subalgebra is regular if and only if for any two
distinct indices $l$ and $l'$, $h_l\neq h_{l'}$.
The Weyl group ${\bf W}(A_{n-1})$ is the permutation group of the $n$ vectors
$\epsilon_l$. The conjugacy classes in ${\bf W}(A_{n-1})$ are in one-to-one
correspondence with the partitions of $n$,
\begin{equation}
(n_1,\ldots, n_s), \quad \sum_{k=1}^s n_k=n,
\label{3.2}\end{equation}
where the $n_k$ ($k=1,\ldots, s$) are non-increasing positive integers
giving the length
of the cycles inside a given conjugacy class.
To describe the action of a representative $w$ of the conjugacy class
associated to the partition (\ref{3.2}), it is useful to re-label the basis
 vectors as follows:
\begin{equation}
\epsilon_{k,{i_k}} := \epsilon_l,
\quad l={(\sum_{m=1}^{k-1} n_m)+i_k},
\quad  k=1,\ldots,s,\  i_k=1,\ldots, n_{k}.
\label{3.3}\end{equation}
The action of $w$ on these basis vectors may be chosen to be
\begin{equation}
w(\epsilon_{k,1})=\epsilon_{k,{n_k}},\quad
 w(\epsilon_{k,{i_k}})=\epsilon_{k,{i_k-1}},\ i_k\neq 1.
\label{3.4}\end{equation}
Since $w$ does
not mix vectors corresponding to different cycles,  one
obtains a basis of
eigenvectors by considering  each cycle separately.
Let us focus our attention on the $k$th cycle of length $n_k$,
and define $\omega_k:=e^{2i\pi\over n_k}$.
The eigenvalues of $w$ on the space spanned by
the vectors  $\epsilon_{k,i_k}$ ($i_k=1,\ldots, n_k$)
are
$(\omega_k)^{j_k}$, $j_k=0,\ldots, n_k-1$,
and the corresponding eigenvectors, denoted as $H_{j_k}(k)$, are
\begin{equation}
H_{j_k}(k)=\sum_{i_k=1}^{n_k} (\omega_k)^{(i_k-1)j_k} \epsilon_{k,i_k}.
\label{3.5}\end{equation}
One can look for a regular eigenvector of $w$ in the form
\begin{equation}
H=\sum_{k=1}^s d_k H_{j_k}(k).
\label{3.6}\end{equation}
The eigenvalues of $w$ on those $H_{j_k}(k)$ for which $d_k\neq 0$
must be equal, and $h_l\neq h_{l'}$ must hold
for any distinct indices
when re-expanding $H$ (\ref{3.6}) in the form (\ref{3.1}).
These conditions lead to the result
summarized in Table 1. Note that $\gcd(p,j)$ denotes the greatest
common divisor of $p$ and $j$, and
in the case  $j=0$ ($\gcd(p,0)=1$)  the condition
$\sum d_k =0$ must be also imposed for the eigenvector to belong to the
Cartan subalgebra of $A_{n-1}$.

\medskip

\begin{center}
\begin{tabular}{c|c|c|c|c|c}\cline{2-5}
& Conjugacy class & Eigenvector & Eigenvalue & Regularity conditions &\\
\cline{2-5} &
$(p,\ldots,p)$,\,
$p\geq1$ & $\sum_{k=1}^s d_k H_j(k)$ &  $\exp({2i\pi j\over p})$ &
${\displaystyle \gcd(p,j)=1 \atop\displaystyle
(d_k)^p\neq (d_{k'})^p,\  d_k\neq 0\  {\rm if}\  p>1}$ & \\ \cline{2-5} &
$( p,\ldots , p , 1)$,\, $p>1$ & $\sum_{k=1}^{s-1}d_k H_j(k)$ & $\exp({2i\pi
j\over p})$ & ${\displaystyle \gcd(p,j)=1 \atop\displaystyle
(d_k)^p\neq (d_{k'})^p,\, d_k\neq 0 }$
 &  \\ \cline{2-5}
\multicolumn{6}{c}{}\\
\multicolumn{6}{c}{Table 1: Regular eigenvectors of
$w\in{\bf W}(A_{n-1})$.}\\
\multicolumn{6}{c}{$H_j(k)=
\sum_{l=1}^p \exp({2\pi i(l-1)j\over p})\epsilon_{(k-1)p+l}\ $
for  $\  0\leq j\leq (p-1)$.}
\end{tabular}\end{center}

\medskip

\subsection{ Regular conjugacy classes in ${\bf W}(D_n)$ }

The Cartan subalgebra of $D_n$  may be identified with the vector space
spanned by $n$ orthonormal vectors $\epsilon_l$, $l=1,\ldots, n$.
The roots of $D_n$ are the vectors
$\pm\epsilon_l\pm\epsilon_{l'}$, $l\neq l'$.
An element $H = \sum_{l=1}^n h_l\epsilon_l$ of the Cartan subalgebra is
regular if and only if for any two distinct indices $l$ and
$l'$, $h_l\neq\pm h_{l'}$.
The Weyl group ${\bf W}(D_n)$ consists of the permutations
of the vectors $\epsilon_l$ and the sign changes of an arbitrary even number
of them \cite{cart}.
A  so called ``signed partition'' of $n$ can be associated to
each conjugacy class,
\begin{equation}
(n_1,\ldots, n_r,\bar n_{r+1},\ldots, \bar n_s), \quad \sum_{k=1}^s n_k=n,
\label{3.7}\end{equation}
where $n_1,\ldots, n_r$ (resp.~$n_{r+1},\ldots, n_s$) is a sequence of
non-increasing positive integers which are the lengths of the positive
(negative)
cycles.
The number of negative cycles $s-r$ is even.
It is shown in \cite{cart}
that a unique conjugacy class in
${\bf W}(D_n)$  is associated to such a signed partition,
except when all cycles are positive of even length, in which
case the same partition corresponds to two distinct conjugacy classes.
To describe the action of a representative $w$ of the conjugacy class
associated to the signed partition (\ref{3.7}),
we follow \cite{leur} and
introduce the adapted basis vectors $\epsilon_{k,i_k}$
($k=1,\ldots, s$, $i_k=1,\ldots,n_k$) similarly to (\ref{3.3}).
The action of $w$ on these basis vectors may be chosen to be:
\begin{equation}
 w(\epsilon_{k,1})=\epsilon_{k,n_k},\quad
 w(\epsilon_{k,i_k})=\epsilon_{k,i_k-1},\  i_k\neq 1,
\quad\hbox{if}\quad 1\leq k\leq r,
\label{3.8}\end{equation}
and
\begin{equation}
w(\epsilon_{k,1})=-\epsilon_{k,n_k},\quad
w(\epsilon_{k,i_k})=\epsilon_{k,i_k-1},\,\ i_k\neq 1,
\quad\hbox{if}\quad r<k\leq s.\label{3.9}\end{equation}
In the case of a signed partition with only positive even cycles, a
representative
$w'$ of the second conjugacy class may be chosen to differ from $w$
(\ref{3.8}) in the
first cycle only, where it contains two sign changes:
\begin{equation}
 w'(\epsilon_{1,1})=-\epsilon_{1,n_1},\quad
w'(\epsilon_{1,2})=-\epsilon_{1,1},\quad
w'(\epsilon_{1,i_1})=\epsilon_{1,i_1-1},\  i_1\neq 1,2.
\label{3.10}\end{equation}
In fact, the conjugacy class of  $w'$  is not regular.
If $H_j(k)$ and $\tilde H_j(k)$ denote
a basis of the eigenvectors of $w$ on
the space spanned by $\epsilon_{k,1},
\ldots, \epsilon_{k,n_k}$ for $k=1,\ldots, r$
 and for $k=r+1,\ldots, s$, respectively,
then the general eigenvector $H$ takes the form
\begin{equation}
H =\sum_{k=1}^r d_k H_{j_k}(k) +
\sum_{k=r+1}^s d_k \tilde H_{j_k}(k),
\label{3.11}\end{equation}
where the eigenvalues of $w$ associated to  the terms with nonzero $d_k$
must be equal.
The eigenvector $H_{j_k}(k)$, with eigenvalue $(\omega_k)^{j_k}$
for $j_k=0,\ldots, n_k-1$, is given in (3.5).
Introduce  the notation $\tilde\omega_k:=e^{2i\pi\over  2n_k}$.
The eigenvector  $\tilde H_{j_k}(k)$,
with eigenvalue  $(\tilde \omega_k)^{2j_k-1}$ for $j_k=1,\ldots, n_k$,
is defined  by
\begin{equation}
\tilde H_j(k)=\sum_{i_k=1}^{n_k}
(\tilde\omega_k)^{(i_k-1)(2j_k-1)} \epsilon_{k,i_k} .
\label{3.11+}\end{equation}
As can be verified by inspecting formula (3.11),
the regular conjugacy classes \cite{Sp} and the corresponding  regular
eigenvector are the ones given in Table 2, where $q$ is an integer.

\medskip

\begin{center}\begin{tabular}{c|c|c|c|c|c}\cline{2-5} &
Conjugacy class & Eigenvector & Eigenvalue & Regularity conditions & \\
\cline{2-5} &
${\displaystyle (p,\ldots,p)\atop\displaystyle
p=2q+\1,\ q\geq 0}$
& $\sum_{k=1}^s d_k H_j(k)$ &
 $\exp({2i\pi j\over p})$ &
${\displaystyle \gcd(p,j)=1\atop\displaystyle
(d_k)^{p}\neq\pm (d_{k'})^{p},\ d_k\neq 0 \ {\rm if}\ p>1}$ & \\
\cline{2-5} &
${\displaystyle (p,\ldots,p,{   1})\atop\displaystyle
p=2q+\1,\ q>0 }$
 & $\sum_{k=1}^{s-1} d_k H_j(k)$ &
$\exp({2i\pi j\over p})$ &
${\displaystyle \gcd(p,j)=1\atop\displaystyle
(d_k)^p\neq\pm (d_{k'})^p,\ d_k\neq 0}$ & \\
\cline{2-5} &
${\displaystyle (\bar p,\ldots ,\bar p)\atop\displaystyle p
\geq \1,\ s=2q,\  q\geq \1}$
& $\sum_{k=1}^s d_k\tilde H_j(k)$ &
 $\exp({2i\pi(2j-1)\over 2p})$ &
${\displaystyle \gcd(p,2j-1)=1
\atop\displaystyle (d_k)^{p}\neq\pm (d_{k'})^{p},\ d_k\neq 0\ {\rm if}\ p>1}$
& \\
\cline{2-5} &
${\displaystyle (\bar p,\ldots ,\bar p, { 1})
\atop\displaystyle p\geq \1,\ s=2q+\1,\  q\geq \1}$
& $\sum_{k=1}^{s-1}d_k\tilde H_j(k)$ &
$\exp({2i\pi(2j-1)\over 2p})$ &
${\displaystyle \gcd(p,2j-1)=1
\atop\displaystyle (d_k)^{p}\neq\pm (d_{k'})^{p},\ d_k\neq 0}$
 &  \\
\cline{2-5} &
${\displaystyle (\bar p,\ldots ,\bar p,\bar {\scriptstyle{1}} ) 
\atop\displaystyle p>\1,\ s=2q,\  q\geq \1}$
& $\sum_{k=1}^{s-1}d_k\tilde H_j(k)$ &
$\exp({2i\pi(2j-1)\over 2p})$ &
${\displaystyle \gcd(p,2j-1)=1
\atop\displaystyle (d_k)^{p}\neq\pm (d_{k'})^{p},\ d_k\neq 0}$
&  \\
 \cline{2-5}
\multicolumn{6}{c}{}\\
\multicolumn{6}{c}{Table 2: Regular eigenvectors of $w\in {\bf W}(D_n)$.}\\
\multicolumn{6}{c}{$H_j(k)=
\sum_{l=1}^{p} \exp({2\pi i(l-1)j\over p})\epsilon_{(k-1)p+l}\ $
for $\ 0\leq j\leq (p-1)$.}\\
\multicolumn{6}{c}{$\ \tilde H_j(k)=
\sum_{l=1}^p \exp({\pi i(l-1)(2j-1)\over p})\epsilon_{(k-1)p+l}\ $
for $\ 1\leq j\leq p$.}
\end{tabular}\end{center}

\subsection{Regular conjugacy classes in ${\bf W}(B_n)\simeq{\bf W}(C_n)$}

We identify the Cartan subalgebra of $B_n$ or $C_n$ with the
 vector space spanned by $n$ orthonormal vectors
$\epsilon_l$, $l=1,\ldots, n$.
The roots of $B_n$ are $\pm\epsilon_l\pm\epsilon_{l'}$,
$l\neq l'$ and $\pm\epsilon_l$.
Those of $C_n$ are $\pm\epsilon_l\pm\epsilon_{l'}$,
$l\neq l'$ and $\pm 2\epsilon_l$.
Thus an
element $H = \sum_{l=1}^n h_l\epsilon_l$ of the Cartan subalgebra is
regular if and only if for any two distinct indices $l$ and $l'$,
$h_l\neq\pm h_{l'}$ and for any $l$, $h_l\neq 0$.
The Weyl groups of $B_n$ and $C_n$ are isomorphic,
they consist of the permutations of the basis vectors $\epsilon_l$
and the sign changes of arbitrary subsets of them.
The conjugacy classes of these groups \cite{cart} are in
 one-to-one correspondence with the signed partitions of $n$:
\begin{equation}
(n_1,\ldots, n_r,\bar n_{r+1},\ldots,\bar n_s),
\quad
 \sum_{k=1}^s n_k=n,
\label{3.27}\end{equation}
where $n_1,\ldots, n_r$ (resp.~$n_{r+1},\ldots, n_s$) is a sequence of
non-increasing positive integers which are the lengths of the positive
(negative)
cycles.
The only difference from the $D_n$ case is that there is now no limitation on
the number of negative cycles. A representative $w$ of the conjugacy class
labelled by
the signed partition (\ref{3.27}) is obtained using the same formulas
(\ref{3.8}), (\ref{3.9}) as in the $D_n$ case.  The
supplementary requirement that for any $l$, $h_l\neq 0$,  simply
prohibits the appearance of
a cycle of length one not contributing to
the eigenvector $H$ in (\ref{3.11}). The result is
 summarized in Table 3, with the same notations as in Table 2.

\bigskip

\begin{center}\begin{tabular}{c|c|c|c|c|c}\cline{2-5}
& Conjugacy class & Eigenvector & Eigenvalue & Regularity conditions & \\
\cline{2-5}
&${\displaystyle
(p,\ldots,p)\atop\displaystyle
p=2q+\1,\ q\geq 0 }$
& $\sum_{k=1}^s d_k H_j(k)$ &
 $\exp({2i\pi j\over p})$ &
${\displaystyle \gcd(p,j)=1\atop\displaystyle
(d_k)^{p}\neq\pm (d_{k'})^{p},\ d_k\neq 0 }$ & \\
\cline{2-5}
& $(\bar p,\ldots ,\bar p)$,\, $p\geq 1$
& $\sum_{k=1}^{s}d_k\tilde H_j(k)$ &
$\exp({2i\pi(2j-1)\over 2p})$ &
${\displaystyle
\gcd(p,2j-1)=1
\atop \displaystyle(d_k)^{p}\neq\pm (d_{k'})^{p},\ d_k\neq 0}$
 &  \\
\cline{2-5}
\multicolumn{6}{c}{}\\
\multicolumn{6}{c}{Table 3: Regular eigenvectors of
$w\in {\bf W}(B_n)\simeq {\bf W}(C_n)$.} \\
\multicolumn{6}{c}{$H_j(k)=
\sum_{l=1}^{p} \exp({2\pi i(l-1)j\over p})\epsilon_{(k-1)p+l}\ $
for $\ 0\leq j\leq (p-1)$.}\\
\multicolumn{6}{c}{$\ \tilde H_j(k)=
\sum_{l=1}^p \exp({\pi i(l-1)(2j-1)\over p})\epsilon_{(k-1)p+l}\ $
for $\ 1\leq j\leq p$.}
\end{tabular}\end{center}

\bigskip

The regular conjugacy
classes in the Weyl group of an  exceptional simple Lie algebra,
and in the group obtained as the extension of the Weyl group  by
the automorphisms of the Dynkin diagram,
 are  also listed in \cite{Sp}.
The classification of regular conjugacy classes
in the extended Weyl groups can be used to find graded regular semisimple
elements in the twisted affine Lie algebras, similarly to the role
of the Weyl group in the non-twisted case to which our attention is
restricted in this paper.

\newpage

\section{ Heisenberg subalgebras
with
graded regular
elements and $sl_2$ embeddings}
\setcounter{equation}{0}

In Section 2
we have seen that the graded Heisenberg subalgebras
of the non-twisted loop algebra
$\ell(\G)$ are classified by the conjugacy classes $[w]$ in ${\bf W}(\G)$,
and the graded regular elements in the Heisenberg subalgebra
$\tilde {\cal H}_{\hat w}\subset \ell(\G)$
arise from the regular eigenvectors of $w\in {\bf W}(\G)$.
For $\G$ a classical Lie algebra,
the conjugacy  classes in ${\bf W}(\G)$
listed in the tables of Section 3  parametrize those Heisenberg
subalgebras that contain graded regular  elements.
In this section we describe  a relationship between these
Heisenberg subalgebras and certain $sl_2$ subalgebras of $\G$.
This relationship consists of two points.
First, in the cases when $\tilde {\cal H}_{\hat w}$
contains a graded regular element, the grading
$d_{N,Y}$ of $\ell(\G)$ induced using
the  appropriately lifted Weyl group  element $\hat w$ in the
construction of Section 2 takes the form
\begin{equation}
d_{N,Y}=\nu d_{m,I_0},
\quad
d_{m,I_0}=m\lambda {d\over d\lambda } +{\rm ad} I_0 ,
\label{4.*}\end{equation}
where
{\em $I_0\in \G$ is the semisimple element of an $sl_2$ subalgebra
$\{ I_-,I_0, I_+\}\subset \G$}
in the normalization
\begin{equation}
[I_0, I_\pm]=\pm I_\pm,
\quad
[I_+,I_-]=2I_0.
\label{4.**}\end{equation}
Here $\nu=1$ or $2$ depending on whether
$I_0$ determines an integral (even)
or a half-integral  $sl_2$ subalgebra of $\G$, and $(m-1)$ is the largest
eigenvalue of ${\rm ad} I_0$ on $\G$.
Second,   for any graded regular element
$\Lambda\in \tilde {\cal H}_{\hat w}$
 of {\em minimal} positive grade, which has the form
\begin{equation}
\Lambda = C_+ +\lambda C_-
\quad\hbox{with some}\quad C_\pm \in \G,
\label{4.***}\end{equation}
we show that $C_+$ {\em is the raising element of an $sl_2$ subalgebra
containing} $I_0$.
That is there exists $I_-\in \G$ such  that (\ref{4.**}) holds with
$I_+:=C_+$.  The  $d_{m,I_0}$ grade of $\Lambda$ is one.
These statements provide a
 generalization of  the well-known relationship between the
principal Heisenberg subalgebra and the principal $sl_2$
embedding,
which underlies the $\W$-algebra structure of the KdV type
hierarchies of Drinfeld and Sokolov \cite{DS}.
In Subsection 4.1  we present  a  convenient  method for constructing
explicit realizations of  the  Heisenberg subalgebras,
which will be used to verify the above statements in Subsection 4.2.

It should be emphasized that the above statements refer to a
particular lift $\hat w$ of $w\in [w]$.
A construction of the appropriate lift  will be given
for any regular conjugacy class $[w]\subset {\bf W}(\G)$.

The correspondence
between certain $sl_2$ subalgebras in $\G$  and certain
conjugacy classes in ${\bf W}(\G)$ has been investigated
in the  mathematics literature from various viewpoints.
The connection of  the above mentioned statements
to related results in \cite{Kost,Sp,cart2}
will be explained in Subsection 4.2.
See also Appendix A.

\subsection{A practical algorithm to construct Heisenberg subalgebras}

Recall that the principal Heisenberg subalgebra of $\ell(\G)$
is associated to the conjugacy class in ${\bf W}(\G)$ consisting of
Coxeter elements \cite{kac}.
The Coxeter class is one of the so called {\em primitive}
conjugacy classes of ${\bf W}(\G)$, which
are  characterized  in \cite{KP,bouw}
by the condition that $\det(1-w)=\det({\cal A})$ for a representative $w$,
where $\cal A$ is the Cartan matrix of $\G$.
In \cite{cart2} the term ``semi-Coxeter'' classes is used to denote
the primitive conjugacy classes.
The most intuitive defining property of
these conjugacy classes is
that they do not possess a representative
contained in  a proper Weyl subgroup of ${\bf W}(\G)$.
The  Weyl subgroups of ${\bf W}(\G)$ are
the Weyl groups  of the regular semisimple subalgebras of $\G$.
For  the algebras $A_{n}$, $B_n$, $C_n$ and $G_2$ the Coxeter
class is the only primitive conjugacy class \cite{cart}.
Concretely, it is the class of the cyclic permutation
$(n+1)$ for ${\bf W}(A_n)$ and that of the negative cycle
$(\bar n)$ for ${\bf W}(B_n)\simeq {\bf W}(C_n)$.
For ${\bf W}(D_n)$ the situation is more
interesting.
The primitive conjugacy classes are those containing two
negative cycles, $(\bar n_1,\bar n_2)$ for any $n_1\geq n_2\geq 1$,
$n_1+n_2=n$, and the Coxeter class is that of $n_2=1$.
The classification of the
conjugacy classes in ${\bf W}(\G)$ described in \cite{cart}
is closely related to the classification of the regular semisimple
subalgebras of $\G$ treated by Dynkin \cite{dyn}.
In fact, it has been
shown\footnote{This is shown in \cite{cart} for any simple Lie algebra
including the exceptional ones.} in \cite{cart}
that each conjugacy class of ${\bf W }(\G)$
can be (in general non-uniquely) represented
by an element $w\in {\bf W}(\G)$
of the product form
\begin{equation}
w=w_1 \cdot w_2 \cdots w_r,
\label{4.1}\end{equation}
where $w_k$ belongs to a primitive conjugacy class
in the Weyl group ${\bf W}(\G_k)$ of the simple factor ${\G}_k$
($k=1,\ldots, r$)
of a regular semisimple subalgebra of $\G$,
\begin{equation}
\G_1 + \G_2 +\cdots +\G_r  \subset \G.
\label{4.2}\end{equation}
The Cartan subalgebra $\H\subset \G$ on which $w$ given in
 (\ref{4.1}) acts is a direct sum
\begin{equation}
\H = \H_1 +\H_2+\cdots + \H_r + \H'\,,
\label{4.3}\end{equation}
where $\H_k$ is a Cartan subalgebra of $\G_k$
and $w$ acts as the identity on the subalgebra
$\H'\subset {\cal H}$ which is orthogonal to ${\cal H}_k$
for $k=1,\ldots,r$ and
satisfies
${\rm rank\,}\G =\left(\sum_k {\rm rank\,}\G_k\right)+{\rm dim\,}\H'$.
For the construction of the corresponding Heisenberg subalgebra,
 one needs to lift $w$ to a finite order inner
automorphism $\hat w$ of $\G$.
Clearly,  the required lift can be taken to have the form
\begin{equation}
\hat w=\exp\left( 2 i\pi {\rm ad} X\right),
\quad
X=X_1+X_2+\cdots +X_r,
\label{4.4}\end{equation}
where $X_k\in \G_k$ defines an
appropriate  lift $\hat w_k$ of $w_k$
to a finite order
inner automorphism of $\G_k$,
\begin{equation}
\hat w_k=\exp\left( 2i\pi {\rm ad} X_k\right),
\quad X_k\in \G_k.
\label{4.5}\end{equation}
Below $X_k$  will be given explicitly.
We are interested in the graded Heisenberg subalgebra
$\tilde\H_{\hat w} =\eta[\ell(\H,\hat w)]\subset \ell(\G)$
associated to $\hat w$.
The twisted homogeneous Heisenberg subalgebra
$\ell(\H,\hat w)\subset \ell(\G,\hat w)$ in (\ref{2.4})
obviously has the direct sum structure
\begin{equation}
\ell(\H,\hat w) = \ell(\H_1,\hat w_1) +
\ell(\H_2,\hat w_2)+\cdots + \ell(\H_r,\hat w_r) + \ell(\H')\,.
\label{4.6}\end{equation}
Using $\hat w$ in (4.7), the ``untwisting'' $\eta$ in (2.3) induces a
corresponding direct
sum structure
\begin{equation}
\tilde \H_{\hat w} = \tilde \H_{1,\hat w_1} +
\tilde \H_{2,\hat w_2}+\cdots + \tilde \H_{r,\hat w_r} +
\ell( \H')\,,
\label{4.7}\end{equation}
where $\tilde \H_{k,\hat w_k}\subset \ell(\G_k)$ is the
Heisenberg subalgebra associated to the finite order
inner automorphism $\hat w_k$ of $\G_k$, and
$\ell( \H') =\H'\otimes {\bf C}[\lambda,\lambda^{-1}]$.
This leads to the two-step
strategy for constructing  the
non-equivalent graded Heisenberg subalgebras of the loop algebras
$\ell(\G)$:
{\it i)} construct all of the  Heisenberg subalgebras
corresponding to the primitive conjugacy classes in the
Weyl groups of the simple Lie algebras;
{\it ii)}
the general case is then obtained by running over the
regular  semisimple subalgebras of $\G$
and inserting the
``primitive Heisenberg subalgebras'' from the first step into the factors.
Although the  presentation of a Heisenberg subalgebra provided
by this scheme is not unique in general,  it is very convenient
in practice.
In particular, this scheme  defines a correspondence
between the Heisenberg subalgebras of $\ell(\G)$
 possessing a graded
regular element and certain regular semisimple subalgebras of the
Lie algebra $\G$.
In the case when $\G$ is a classical Lie algebra,
the correspondence is summarized in Table 4.

The notations used in Table 4  are as follows.
A simple factor  $\G_k$ appearing in the regular reductive subalgebra
in the third column of the table  represents the
Coxeter class of ${\bf W}(\G_k)$ as well as the
 principal Heisenberg subalgebra of $\ell(\G_k)$.
Concerning the primitive conjugacy classes in the $D_n$ case,
recall from Table 2 that in addition to the Coxeter class
the other  ``extreme case''  $(\bar p, \bar p)$ also
admits a  regular eigenvector for $n=2p$.
The term $\bar D_{2p}$ in Table 4 represents
the conjugacy class
$(\bar p,\bar p)$ of ${\bf W}(D_{2p})$ and the respective
non-principal primitive Heisenberg subalgebra of $\ell(D_{2p})$.
The term $\H'_k$ denotes a Cartan piece of dimension
$k$,  and its presence means that
the subspace  $\ell(\H'_k)$ of the homogeneous Heisenberg
subalgebra $\ell(\H)\subset \ell(\G)$ is contained in $\tilde \H_{\hat w}$.
Since  explicit realizations of the
principal Heisenberg subalgebra of $\ell(\G)$ are known
for every simple Lie algebra,
an explicit realization of any
Heisenberg subalgebra appearing in Table 4 may be obtained
if one constructs one for the primitive case $\bar D_{2p}$.
This will be provided in the next subsection.

\bigskip

\begin{center}\begin{tabular}{c|c|c|c|c|c}\cline{2-5} &
Algebra & Conjugacy class  & Regular
subalgebra & ord($\hat w$)&\\
\cline{2-5} \cline{2-5} &
$A_{ps-1}$
& $(p,\ldots,p)$ & $A_{p-1}+\ldots+A_{p-1}+{\cal H}'_{s-1}$ & $p$ &\\
\cline{2-5} &
$A_{p(s-1)}$
 & $(p,\ldots,p,1)$
 & $A_{p-1}+\ldots+A_{p-1}+{\cal H}'_{s-1}$&$\mbox{gcd}(2,p)p$\\
\cline{2-5} \cline{2-5} &
$D_{ps}$
& $(p,\ldots,p),\,p\,\,\mbox{odd}$ &
$A_{p-1}+\ldots+A_{p-1}+{\cal H}'_s$  &$p$& \\
\cline{2-5} &
$D_{p(s-1) +1}$
& $(p,\ldots,p,1),\, p\,\,\mbox{odd}$  &
$A_{p-1}+\ldots+A_{p-1}+{\cal H}'_s$ & $p$
 &  \\
\cline{2-5} &
$D_{ps}$
& $(\bar p,\ldots ,\bar p),\, s
\,\,\mbox{even}$  &
$ {\bar  D}_{2p}+\ldots+ \bar D_{2p}$ & $2p$
&  \\
 \cline{2-5} &
$D_{p(s-1)+1}$
& $(\bar p,\ldots ,\bar p,\bar 1),\, s\,\, \mbox{even}$  &
$ \bar D_{2p}+\ldots+ \bar D_{2p}+D_{p+1}$ &$2p$
&  \\
 \cline{2-5} &
$D_{p(s-1)+1}$
& $(\bar p,\ldots ,\bar p,1),\, s\,\,\mbox{odd}$ &
$ \bar D_{2p}+\ldots+ \bar D_{2p}+{\cal H}'_1$  &$2p$
&  \\
 \cline{2-5}  \cline{2-5} &
$B_{ps}$
& $(p,\ldots,p),\, p\,\,\mbox{odd}$ &
$A_{p-1}+\ldots+A_{p-1}+{\cal H}'_s$ & $p$
& \\
\cline{2-5} &
$B_{ps}$
& $(\bar p,\ldots ,\bar p),\, s\,\,\mbox{even}$  &
$ \bar D_{2p}+\ldots+\bar D_{2p}$ &$2p$
&  \\
 \cline{2-5} &
$B_{ps}$
& $(\bar p,\ldots ,\bar p),\, s\,\,\mbox{odd}$  &
$\bar  D_{2p}+\ldots+\bar D_{2p}+B_p$ &$2p$
&  \\
 \cline{2-5}  \cline{2-5} &
$C_{ps}$
& $(p,\ldots,p),\, p\,\,\mbox{odd}$  &
$A_{p-1}+\ldots+A_{p-1}+{\cal H}'_s$ & $p$
& \\
\cline{2-5} &
$C_{ps}$
& $(\bar p,\ldots ,\bar p)$  &
$C_p+\ldots+C_p$ &$2p$
&  \\
 \cline{2-5}
\multicolumn{6}{c}{}\\
\multicolumn{6}{c}{Table 4: Heisenberg subalgebras possessing a graded
regular element.}\\
\multicolumn{6}{c}{Here $s$ is the number of cycles in the partition,
$p$ is a positive integer and $A_0=\emptyset$.}
\end{tabular}\end{center}

\medskip

\newpage

\subsection{A connection to $sl_2$ embeddings}

For any simple Lie algebra $\G$, there exists
a celebrated relationship \cite{Kost}
between the Coxeter class
of ${\bf W}(\G)$ and the conjugacy class of the
principal $sl_2$ subalgebra of $\G$,
whose essence is that the lift of a Coxeter element
$w_c\in {\bf W}(\G)$
may be chosen as
\begin{equation}
\hat w_c=\exp\left( 2i\pi {{\rm ad}I_0 \over N_c}\right),
\label{4.11}\end{equation}
where $N_c$ is the Coxeter number and $I_0$ is
the semisimple element of a  principal $sl_2$ subalgebra of $\G$.
This means that there exists $I_\pm\in \G$ so that
\begin{equation}
[I_0, I_\pm]=\pm I_\pm,
\quad
[I_+,I_-]=2I_0,
\label{4.12}\end{equation}
and $I_0$ has the form $I_0={1\over 2} \sum_{\alpha>0} H_{\alpha}$,
where the $H_\alpha\in \bar{\cal H}$ are the Cartan generators
associated to  a system of  positive roots $\alpha >0$
with respect to a Cartan
subalgebra  $\bar {\cal H}\subset \G$.
The Cartan subalgebra $\bar {\cal H}\subset \G$
is said to be ``in apposition'' to
the Cartan subalgebra ${\cal H}\subset \G$ on which $w_c$
acts\footnote{Equivalently, if the principal $sl_2$ generator $I_0$
is taken from ${\cal H}$ then $\hat w_c$ defined by (4.11)
acts as a Coxeter element on the Cartan subalgebra in apposition
$\bar {\cal H}$, which may be defined as the centralizer of
an element $(I_++C_-)\in \G$, where $C_-\neq 0$ is chosen in such
a way that $[I_0,C_-]=-(N_c-1)C_-$.}.
A consequence of this is that the grading of $\ell(\G)$ induced
by its isomorphism with $\ell(\G, \hat w_c)$ is the
principal grading defined by
\begin{equation}
d_{N_c,I_0}=N_c \lambda {d\over d\lambda} +{\rm ad} I_0.
\label{4.13}\end{equation}
Furthermore, decomposing $\G$ as
\begin{equation}
\G=\G^{I_0}_{<0} + \G^{I_0}_0 +\G^{I_0}_{>0}
\label{4.14}\end{equation}
using the (principal) grading of $\G$ defined by ${\rm ad} I_0$,
the grade $1$  regular element $\Lambda$ of the principal Heisenberg
subalgebra $\tilde \H_{\hat w_c}$ takes the form
\begin{equation}
\Lambda = C_+ +\lambda C_-,
\quad
C_\pm \in \G,
\quad\hbox{with}\quad C_+ = I_+,
\label{4.15}\end{equation}
i.e., the $sl_2$ subalgebra of $\G$ defined
by the nilpotent element $C_+\in \G$ through the Jacobson-Morozov
theorem \cite{jac}
is the same $sl_2$ that enters the grading (4.13).
Note also that
\begin{equation}
[C_-, \G^{I_0}_{<0}]=\{0\}.
\label{4.16}\end{equation}
The relations expressed by formulas (4.11), (4.15), (4.16)
play an important role in the Drinfeld-Sokolov construction
of KdV type hierarchies and
we wish to show that they
generalize to all cases given in Table 4,
 for which a graded regular element
exists in the Heisenberg subalgebra.
(The case of the homogeneous Heisenberg subalgebra is
related to the trivial, identically zero, $sl_2$ embedding and is excluded
in what follows.)
We need to deal with  the $\bar D_{2p}$
case first, since it occurs  as a ``building block''
in Table 4.

In order to take care of  the $\bar D_{2p}$ case, we make use
of a result of \cite{leur} on the lift of
a Weyl group element $w_{(\bar n_1,\bar n_2)}\in {\bf W}(D_n)$
belonging to the conjugacy class
$(\bar n_1,\bar n_2)$.
In Section 2.6 of \cite{leur},
a lift $\hat w_{(n_1,n_2)}$   conjugate to
\begin{equation}
\hat \tau_{(\bar n_1,\bar n_2)}:=\exp\left(2i\pi {\rm ad} K/N\right),
\label{4.17}\end{equation}
where $N={\rm lcm}(2n_1,2n_2)$ is the order of
$\hat \tau_{(\bar n_1,\bar n_2)}$ and
\begin{equation}
K={N\over 2n_1}\sum_{k=1}^{n_1} (n_1-k+1) \epsilon_k
+{N\over 2n_2}\sum_{k=1}^{n_2} (n_2-k) \epsilon_{n_1+k},
\label{4.18}\end{equation}
was constructed  for any $n_1+n_2=n$.
We  observe that $K$ is the semisimple element of an $sl_2$ subalgebra
of $D_n$ in the Coxeter case $n_2=1$ and in the case $n_1=n_2$,
and is not proportional to such an element
in the other cases.
This is most easily seen from the spectrum of the matrix
$K$ in the defining $2n$ dimensional representation of $D_n$,
taking into account
that  $\epsilon_k$ ($k=1,\ldots, n$) contains two
non-zero entries, $\pm 1$, when
diagonalized.
For $n_1=n_2=p$, this explicit form of $K$ also implies that
the $4p$ dimensional
vector representation of $D_{2p}$ decomposes under the
$sl_2$ subalgebra containing $K$ according to
\begin{equation}
4p=(2p+1)+(2p-1).
\label{4.19}\end{equation}
According to Dynkin \cite{dyn},
this is one of the singular $sl_2$ subalgebras
(``$S$-subalgebras'') in $D_{2p}$.
(Note that the singular $sl_2$ subalgebras of \cite{dyn} are called
semi-regular $sl_2$ subalgebras and the principal $sl_2$
is called the regular $sl_2$ in some of the literature.)
It is interesting  that the number of conjugacy classes
of singular $sl_2$ subalgebras in $D_n$ is actually equal to the number
of primitive conjugacy classes in ${\bf W}(D_n)$, but the above
lift of $w_{(\bar n_1,\bar n_2)}$ corresponds to an $sl_2$
embedding only in the  cases when
$w_{(\bar n_1,\bar n_2)}$ admits a regular eigenvector.

It follows from the above that $\hat \tau_{(\bar p,\bar p)}$ in (4.17)
may be used in the
construction of the sought after Heisenberg subalgebra,
denoted as $\tilde \H_{(\bar p,\bar p)}$, where $K$
is the semisimple $sl_2$ generator corresponding
to the decomposition (4.19) of the defining representation
of $D_{2p}$ (which defines it up to conjugation).
It is convenient to realize $D_{2p}$ as
the subalgebra of $gl_{4p}$ consisting of the matrices $A$
subject to $A^t \eta + \eta A=0$ with the
$4p\times 4p$ matrix $\eta$ given by
\begin{equation}
\eta:=\sum_{k=1}^{2p+1} e_{k, 2p+2-k} + \sum_{k=1}^{2p-1}
e_{2p+1+k, 4p+1-k},
\label{4.20}\end{equation}
where $e_{i,j}$ is the usual elementary
matrix with a single non-zero entry $1$ at the $ij$ position,
and to realize the $sl_2$ generator $K$ as
\begin{equation}
K={\rm diag}\left(p, \ldots,0,\ldots,  -p,
(p-1), \ldots,0,\ldots, -(p-1)\right).
\label{4.21}\end{equation}
 The  appropriate grading of $\ell(D_{2p})$
is given by
\begin{equation}
d_{2p,K}=2p\lambda {d\over d\lambda} +{\rm ad}K.
\label{4.22}\end{equation}
Note also from Table 2 that the grade $q$ subspace
of $\tilde \H_{(\bar p,\bar p)}$
 must be of dimension $2$ if
$q=1, 3,\ldots, (2p-1)$ modulo $2p$, and is otherwise empty.
Let us now introduce  the  matrices $H_{1,1}$ and $H_{1,2}$ in $D_{2p}$,
\begin{eqnarray}
&&H_{1,1}:=\sum_{k=1}^p a_k e_{k,k+1} -\sum_{k=1}^{p}a_{p+1-k} e_{p+k,p+k+1}
+a_{p+1}(e_{2p,1} - e_{2p+1,2}), \label{4.23}\\
&&H_{1,2}:=\sum_{k=1}^{p-1} b_k e_{2p+k+1,2p+k+2} -
\sum_{k=1}^{p-1} b_{p-k} e_{3p+k,3p+k+1}+b_p a_1(e_{4p,2p+1} -e_{1,2p+2})
+\nonumber\\
&&\phantom{ H_{1,2}:=}b_p a_{p+1} (e_{4p,1}-e_{2p+1,2p+2}),
\nonumber \end{eqnarray}
where $a_1,\ldots,a_{p+1},b_1,\ldots,b_{p}\in {\bf C}$ are
arbitrarily chosen {\it non-zero} parameters.
We also need their matrix powers
\begin{equation}
H_{j,k}:=\left(H_{1,k}\right)^{2j-1}
\quad\hbox{for}\quad
j=1,2,\ldots,p,\quad k=1,2.
\label{4.24}\end{equation}
It can be checked  that these $2p$ matrices commute
and span a Cartan subalgebra of $D_{2p}$ for generic choice
of the parameters.
We denote this Cartan subalgebra as $\H_{(\bar p,\bar p)}$.
The point is that $\H_{(\bar p,\bar p)}\subset D_{2p}$
is invariant under the automorphism given in (4.17),
and $H_{j,k}$ is  the corresponding basis of eigenvectors:
\begin{equation}
\hat \tau_{(\bar p,\bar p)}\left(H_{j,k}\right) =
(\omega)^{2j-1} H_{j,k},
\quad
\omega= \exp\left({2i\pi \over N}\right),
\quad N=2p.
\label{4.25}\end{equation}
This implies that
$\hat \tau_{(\bar p,\bar p)}$ acts on the Cartan subalgebra
$\H_{(\bar p,\bar p)}$ as a representative
of the conjugacy class $(\bar p,\bar p) \subset {\bf W}(D_{2p})$.
Performing the ``untwisting'' described in Section 2 is straightforward,
and we get the Heisenberg subalgebra
$\tilde \H_{(\bar p,\bar p)}\subset \ell(D_{2p})$
as the span of the following graded basis:
\begin{equation}
\lambda^m \left(\Lambda_{1,k}\right)^{2j-1},
\quad\hbox{for}\quad
m\in {\bf Z},\  j=1,2,\ldots,p,\  k=1,2,
\label{4.26}\end{equation}
where
\begin{eqnarray}
&&\Lambda_{1,1}:=\sum_{k=1}^p a_k e_{k,k+1} -\sum_{k=1}^{p}a_{p+1-k}
e_{p+k,p+k+1}
+\lambda a_{p+1}(e_{2p,1} - e_{2p+1,2}), \label{4.27}\\
&&\Lambda_{1,2}:=\sum_{k=1}^{p-1} b_k e_{2p+k+1,2p+k+2} -
\sum_{k=1}^{p-1} b_{p-k} e_{3p+k,3p+k+1}+b_p a_1(e_{4p,2p+1} -e_{1,2p+2})
+\nonumber\\
&&\phantom{ \Lambda_{1,2}:=} \lambda b_p a_{p+1} (e_{4p,1}-e_{2p+1,2p+2}).
\nonumber \end{eqnarray}
The basis vector in (4.26)
has grade $(2j-1)+2mp$
with respect to the grading $d_{2p,K}$.
This construction of $\tilde \H_{(\bar p,\bar p)}$
was inspired by an analogous construction in \cite{leur}.
A grade 1 regular element
$\Lambda \in \tilde \H_{(\bar p,\bar p)}$ will be a linear
combination
\begin{equation}
\Lambda=d_1 \Lambda_{1,1} + d_2 \Lambda_{1,2}
\label{4.28}\end{equation}
with generic non-zero coefficients $d_1, d_2$.
Writing $\Lambda$ in the form $\Lambda=C_+ +\lambda C_-$,
$C_+$ has grade $1$ and $C_-$ has grade $-(2p-1)$
with respect to ${\rm ad} K$.
We wish to show that   $K$ and $C_+$
are contained in the same $sl_2$ subalgebra of $D_{2p}$,
i.e., that the commutation relations given in (\ref{4.12}) hold
with $I_0:=K$, $I_+:=C_+$ and some $I_-\in D_{2p}$, analogously
to the principal case.

We need to present an auxiliary result at this point.
Consider  a  regular semisimple element
$\Lambda =(C_+ + \lambda C_-)\in \ell(\G)$,
with some  $C_\pm \in \G$,
having  definite grade  with respect to a grading  operator
$d_{N,K}=N\lambda {d\over d\lambda} + {\rm ad} K$.
Suppose that
\begin{equation}
[C_-, \G^K_{<0}]=\{ 0\},
\label{4.29}\end{equation}
where $\G=\G^K_{>0}+\G^K_0+\G^K_{<0}$ is  the decomposition defined by
means of the eigenvalues of ${\rm ad} K$.
Then the following ``non-degeneracy relation''
\begin{equation}
{\rm Ker}\Bigl({\rm ad\,} C_+\Bigr) \cap \G^K_{<0} = \{ 0\}
\label{4.30}\end{equation}
is satisfied.
Indeed, if one could find an element $v\in \G^K_{<0}$ for which $[C_+,v]=0$,
then $[\Lambda, v]=0$ would also hold because of (\ref{4.29}).
Clearly,  ${\rm Ker}\left( {\rm ad } \Lambda\right)\subset
\ell(\G)$ can contain only
semisimple  elements of $\G\subset \ell(\G)$,
but any $v\in \G^K_{<0}$ is a nilpotent element.
This contradiction proves (\ref{4.30}).

The above argument applies to $\Lambda$ in (\ref{4.28}),
since (\ref{4.29}) follows from  the fact that the grade of $C_-$ is
the smallest  eigenvalue of ${\rm ad} K$ on $\G=D_{2p}$.
A consequence of the non-degeneracy relation (\ref{4.30}) is the
equality ${\rm dim\,}[C_+, \G^K_{-1}] = {\rm dim\,} \G^K_{-1}$.
This implies the existence of $I_-\in \G^K_{-1}$ for
which $[C_+,I_-]=K$, since in our case
${\rm dim\,} \G^K_{-1}={\rm dim\,}\G^K_0$ holds as is easily verified
using the explicit formula (\ref{4.21}) of the grading operator $K$.
The set $\{ I_-, I_0:=K,  I_+:=C_+\}$
spans the required $sl_2$ subalgebra.
This settles  the $\bar D_{2p}$ case.

\medskip

Turning now to the general case,
we first rewrite the lift $\hat w$ in (\ref{4.4}) as
\begin{equation}
\hat w = \exp\left( 2i\pi {{\rm ad} Y \over N }\right),
\label{4.31}\end{equation}
where
\begin{equation}
Y=NX={N\over N_1} Y_1 + {N\over N_2} Y_2 +\cdots + {N\over N_r} Y_r.
\label{4.32}\end{equation}
Here $N$ is the order of $\hat w$, $N_k$ is the order of $\hat w_k$
when acting on $\G_k$, $Y_k=N_k X_k$ in (\ref{4.5}).
The grading of $\ell(\G)$ corresponding to $\hat w$ is defined
by the operator $d_{N,Y}$,
\begin{equation}
d_{N,Y}
=N\lambda {d\over d\lambda} + {\rm ad\,}Y.
\label{4.43}\end{equation}
When restricted to the subalgebra $\ell(\G_k)$, this
grading satisfies
\begin{equation}
d_{N,Y}\vert_{\ell(\G_k)}={N\over N_k} d_{N_k,Y_k},
\quad
d_{N_k,Y_k}=N_k \lambda {d\over d \lambda} +{\rm ad\,} Y_k ,
\label{4.34}\end{equation}
where $d_{N_k, Y_k}$ gives the grading of
$\ell(\G_k)$ induced by the isomorphism
$\ell(\G_k)\simeq \ell(\G_k, \hat w_k)$.
Using the lifts of the regular primitive
Weyl transformations given in (4.11) and in (4.17),
$Y_k$ is the semisimple element
of an $sl_2$ subalgebra of $\G_k$ with the same normalization as
$I_0$ in (\ref{4.12}).
Hence it follows from (\ref{4.32}) that $Y$ is proportional to
the semisimple element of an $sl_2$ subalgebra of $\G$ if and only if
\begin{equation}
N_i=N_j
\quad\hbox{for any}\quad
i\neq j.
\label{4.35}\end{equation}
Inspection shows that (\ref{4.35}) is satisfied for all cases
in Table 4, and therefore
\begin{equation}
Y= {N\over N_1}( Y_1 + Y_2 +\cdots + Y_r),
\label{4.36}\end{equation}
where ${N\over N_1}$ turns out
to be $1$ or $2$ depending on whether
$(Y_1+Y_2+\cdots+Y_r)$ defines an integral or a half-integral
$sl_2$ subalgebra of $\G$, i.e., whether the grading of $\G$
defined by this element is integral or half-integral.
In fact, the $sl_2$ embedding is  an integral one in all cases
in Table 4 except the case $(p,\ldots,p,1)$ with $p$ {\it even}
for $\G=A_{p(s-1)}$.
One also sees that any graded regular semisimple element
$\Lambda\in \tilde {\cal H}_{\hat w}$
of minimal positive grade (${N/N_1}$)  has the form
$\Lambda =C_+ +\lambda C_-$, where
$I_+:=C_+$ is contained  in  an $sl_2$ subalgebra whose semisimple
element is $I_0:={N_1\over N}Y$ given by (\ref{4.32}).
This is a consequence
of what we know about the  principal and $\bar D_{2p}$ cases,
simply because such a $\Lambda$
is a linear combination of respective graded regular elements from the
Heisenberg subalgebras $\tilde \H_{\hat w_k}\subset
\ell(\G_k)$ in (\ref{4.7}).
With respect to the grading of $\G$  defined by ${\rm ad} I_0$,
the non-degeneracy relation
\begin{equation}
{\rm Ker}\Bigl({\rm ad\,} I_+\Bigr) \cap \G^{I_0}_{<0} = \{ 0\}
\label{4.37}\end{equation}
then follows from the $sl_2$  structure.
Inspection  shows that $[C_-, \G^{I_0}_{<0}]=\{0\}$ is also
satisfied in each case, since
$C_-$ is an eigenvector of ${\rm ad} I_0$
associated to the smallest eigenvalue.

\medskip

Let us summarize the  results obtained in this section.
For $\G$ a classical Lie algebra,
we  verified the following  connection  between regular conjugacy
classes in ${\bf W}(\G)$,
with  graded regular elements in
the associated Heisenberg subalgebra
$\tilde {\cal H}_{\hat w}\subset \ell(\G)$,
and $sl_2$ subalgebras in $\G$.
For any regular conjugacy class $[w]\subset {\bf W}(\G)$,
the appropriately lifted Weyl transformation takes the
form $\hat w=\exp\left(2i\pi {\rm ad}Y/N\right)$ in (4.31), where
$Y=\nu I_0$ with $I_0$ being the  semisimple generator
of an $sl_2$ subalgebra of $\G$ and $\nu=1$ or $\nu=2$
so that ${\rm ad } Y$ has integral eigenvalues.
The largest eigenvalue of ${\rm ad} Y$ is  $(N-\nu)$, where $N$
is the order of $\hat w$ and $m=N/\nu$ is the order of $w\in [w]$.
The smallest positive $d_{m,I_0}$ grade
for which a graded regular element
$\Lambda \in\tilde {\cal H}_{\hat w}$ exists is one,
and any grade one regular element has the form
$\Lambda=(C_++\lambda C_-)$, where
$C_+$ is  included in  an $sl_2$ subalgebra containing
also $I_0$.  The eigenvalue of ${\rm ad} I_0$ is minimal on $C_-$.
Of course  $\hat w$  acts as the Weyl
transformation $w$ on the Cartan subalgebra defined by the centralizer
of its regular semisimple eigenvector given by
 $H:=\Lambda(\lambda=1)=(C_+ + C_-)\in \G$.

If $w$ in (4.4)  is a Coxeter element in  a regular semisimple
subalgebra of $\G$,
the above  results followed  from the result
of Kostant \cite{Kost} on the connection
between  the Coxeter class  and the principal  $sl_2$
given by formula (4.11).
The case of the regular semi-Coxeter
conjugacy class $(\bar p, \bar p)\subset
{\bf W}(D_{2p})$ was  dealt with  by inspecting the lift
found  in \cite{leur}.

We wish to  note that in  \cite{Sp}
the result of Kostant \cite{Kost} was
generalized to give a similar connection  between certain regular conjugacy
classes in ${\bf W}(\G)$ and those special integral $sl_2$ subalgebras
of $\G$ for which the decomposition of $\G$ into $sl_2$ irreducible
representations
contains no singlets and only one triplet.
In addition to the principal $sl_2$, such $sl_2$ subalgebras
exist only in the exceptional Lie algebras as listed in
\cite{Sp}
\footnote{The $sl_2$ subalgebra of $G_2$ appearing  in Table 11 of \cite{Sp}
has in fact $3$ triplets and not $1$, but the claims  are still
valid for this $sl_2$ as is easily seen using
that  it is actually the principal $sl_2$ inside
the regular  $A_2\subset G_2$.}.  See also Appendix A.

In passing, we also wish to mention
the  correspondence found in \cite{cart2}
between the conjugacy classes of arbitrary  singular (semi-regular)
$sl_2$ subalgebras
in $\G$ \cite{dyn} and
a subset of the primitive (semi-Coxeter) conjugacy classes
in ${\bf W}(\G)$.
This is given in terms of an injective mapping from
the set of singular $sl_2$ subalgebras into the set of
primitive conjugacy classes, which is defined
by the coincidence
 of the so called ``Carter diagrams'' \cite{cart}
associated  to the conjugacy classes in ${\bf W}(\G)$
and to the $sl_2$  subalgebras in $\G$.
On the overlap of their ranges of applicability,
the ``Kostant type'' correspondence discussed in \cite{Sp},
and  here for $D_{2p}$,  and the one in  \cite{cart2} are consistent.
It is not clear whether the result of \cite{cart2}
has any significance  for the theory of integrable hierarchies.

\newpage

\section{Applications to KdV type systems}
\setcounter{equation}{0}

Now we turn to the application of the results collected
in the previous sections  to the construction of integrable
hierarchies.
For $\G$ a simple Lie algebra,
fix  a grade one  regular semisimple element $\Lambda$ from a
Heisenberg subalgebra $\tilde {\cal H}_{\hat w}\subset \ell(\G)$.
Suppose  that $\Lambda$  has  the form
\begin{equation}
 \Lambda = I_+ + \lambda C_- ,
\label{5.1}\end{equation}
where $I_+$ belongs to the $sl_2$ subalgebra
$\{ I_-, I_0, I_+\}\subset \G$
for which $d_{m,I_0}$ in (4.1) defines the grading of $\ell(\G)$.
Suppose also that
\begin{equation}
[C_-, \G_{<0}]=\{0\}
\qquad \hbox{with}\qquad
\G_{<0}=\sum_{k<0} {\cal G}_k ,
\label{5.2}\end{equation}
where $\G_k$, denoted in Section 4 as $\G_k^{I_0}$,
is  the eigensubspace of ${\rm ad} I_0$ with eigenvalue  $k$.

As we have seen,
for $\G$ a classical Lie algebra the relations
in (5.1) and (5.2) are ensured by  using  the
lift $\hat w$ given in (4.31) for an arbitrary regular
conjugacy class $[w]\subset {\bf W}(\G)$.
For the exceptional Lie algebras these relations may be
assumed in connection with many  regular conjugacy classes in
the Weyl group, which include for example all regular conjugacy
classes in ${\bf W}(G_2)$ and all of the
regular primitive conjugacy classes.
It appears that in ${\bf W}(F_4)$, ${\bf W}(E_{6,7,8})$
there  exist  some regular conjugacy classes
for which (5.2) cannot be satisfied;  see Appendix A.

Let us recall \cite{bais,rep} that one may associate a
 ``classical $\W$-algebra'' to any
$sl_2$ subalgebra of $\G$  by a generalization of the Hamiltonian reduction
used by Drinfeld and Sokolov to obtain the
second (Gelfand-Dickey) Poisson bracket of their KdV type hierarchies.
In Subsection 5.1 we  show that if the $sl_2$ subalgebra is related to
a grade one  regular  semisimple element  $\Lambda$ in the above way,
which specifies a (small) subset of the
non-equivalent $sl_2$ subalgebras of $\G$,
then it is possible  to obtain a KdV type hierarchy from
Hamiltonian reduction whose second Poisson bracket is
the $\W$-algebra defined by the  $sl_2$-subalgebra.
Subsection 5.2 is devoted  to the concrete
description of some of the systems that may be obtained from
this  approach.
We  analyze the cases when $\G$ is a classical Lie
algebra of $B$, $C$ or $D$ type and the regular reductive subalgebra
appearing in the third column  of Table 4 contains only $A$ or $C$ type
simple factors.
The resulting  generalized KdV systems turn out to be
discrete reductions of the systems associated to $gl_n$
having the Gelfand-Dickey type Lax operators in (1.3) and (1.4).
That is  the  Lax operators of the resulting systems are of the
form (1.3) or (1.4) subject to certain  extra symmetry conditions,
very much like  the well-known  principal case \cite{DS}
 for the Lie algebra
 $C_p$, where  the
Lax operator is of the form (1.1) with $n=2p$
subject to the self-adjointness condition $L^\dagger =L$.

\subsection{KdV systems associated to grade one regular elements}

The following construction is a straightforward generalization of
that in \cite{DS}, and can be also viewed as a special case  of the
more general construction given in \cite{McI,Prin1,Prin2}.

After fixing   a grade one  regular semisimple element
$\Lambda \in \ell(\G)$ subject to (5.1), (5.2),  consider
the  manifold  $\M$ consisting of  first order differential operators,
\begin{equation}
\M:= \left\{\, {\cal L}= \pa + J + \lambda C_-\, \vert\,
J\in \cinf\left(S^1,{\cal G}\right) \,\right\}.
\label{5.3}\end{equation}
The manifold $\M$ is the phase space of an infinite collection
of bi-Hamiltonian systems.
The two compatible Poisson brackets (PBs) are given as follows.
The ``second'' PB
is given by the affine current algebra structure,
\begin{equation}
\{ f, h \}_2(J) = \int_{S^1}\tr\left(
J\left[{\delta f\over \delta J}, {\delta h \over \delta J}\right]
+\left({\delta f\over \delta J}\right)' {\delta h\over \delta J}\right),
 \label{5.4}\end{equation}
and the ``first'' PB is given by
\begin{equation}
\{ f, h \}_1(J) = -\int_{S^1}\tr
C_-\left[{\delta f\over \delta J}, {\delta h \over \delta J}\right],
\label{5.5}\end{equation}
for $f$, $h$ smooth functions on ${\cal M}$.
The  Hamiltonians of interest  are generated
by the invariants  (``eigenvalues'') of the monodromy matrix
$T(J,\lambda)$ of ${\cal L}$.
The corresponding Hamiltonian flows commute as a special case
of the Adler-Kostant-Symes
construction  and are bi-Hamiltonian (see e.g.~\cite{RSTS2}) .
The Hamiltonians given by the
monodromy invariants are non-local functionals of $J$ in general.
Using that  $C_-$ in (5.3) is related to the
regular semisimple element $\Lambda$ according to  (\ref{5.1}), we can
perform a symmetry reduction of the system on $\M$ leading to
a {\it local} hierarchy.

Let $G$ be a connected Lie group corresponding to $\G$.
Define the  subgroup  ${\rm Stab}(C_-)$ of $G$
by $gC_-g^{-1}=C_-$ for $g\in {\rm Stab}(C_-)$.
Denote  the group of smooth loops
based on ${\rm Stab}(C_-)$ as
$\widetilde{{\rm Stab}}(C_- ):=\cinf\left(S^1, {\rm Stab}(C_-)\right)$.
The possibility for reduction rests upon  the
fact  that there is a Poisson action
(meaning that it leaves the PBs unchanged)
of $\widetilde{{\rm Stab}}(C_-)$  on ${\cal M}$ given by
\begin{equation}
(\pa + J + \lambda C_-) \mapsto g (\pa  + J + \lambda C_-)g^{-1}
= g (  \pa + J)g^{-1} + \lambda C_-,
\quad
\forall\, g\in \widetilde{{\rm Stab}}(C_-),
\label{5.6}\end{equation}
which leaves the monodromy invariants unchanged.
For present purposes we consider
reduction based on the subgroup ${\cal N}$ of $\widetilde{{\rm Stab}}(C_-)$
whose  Lie algebra is $C^\infty\left(S^1, \G_{<0}\right)$.
The  reduction is defined by first imposing constraints on
${\cal M}$ so that the constrained submanifold $\M_c\subset \M$ is
\begin{equation}
\M_c:= \left\{ {\cal L} = \pa + j + \Lambda\,\vert\,
 j\in \cinf\left(S^1,\G_{<1}\right)
 \,\right\},
\qquad\left( \G_{<1}=\sum_{k<1} {\cal G}_k\right).
\label{5.7}\end{equation}
That is the constraints defining ${\cal M}_c \subset {\cal M}$
restrict $J$ to have the form $J=(j+I_+)$ with $I_+$ in (5.1).
The second step of the reduction is to factorize
${\cal M}_c$ by the group ${\cal N}$ of
``gauge transformations'' acting according to
\begin{equation}
e^f: {\cal L} \mapsto e^f{\cal L}e^{-f},
\qquad
\forall\,e^f\in\N,  \quad\hbox{with}
\quad f\in \cinf\left(S^1,{\cal G}_{<0}\right).
\label{5.8}\end{equation}

Standard arguments show that the compatible PBs on ${\cal M}$
induce compatible PBs on the space of
gauge invariant functions on ${\cal M}_c$, identified as
the space of functions on the reduced space
${\cal M}_{\rm red}={\cal M}_c/{\cal N}$.
Thanks to the non-degeneracy relation in (\ref{4.37}),
the  gauge  fixing procedure of \cite{DS} is applicable to
obtain a  basis of the gauge invariant differential polynomials
on ${\cal M}_c$, which may be used as coordinate functions on $\M_c/\N$.
The gauges resulting from this procedure are often called
``DS gauges''  (see e.g.~\cite{rep}).
A particular DS gauge is the so called lowest weight gauge \cite{bal},
whose gauge section ${\cal M}_{\rm l.w.}\subset {\cal M}_c$
is defined as
\begin{equation}
{\cal M}_{\rm l.w.}:=\left\{\,
{\cal L}=\pa + j_{\rm l.w.} + \Lambda \,\vert\,
j_{\rm l.w.}\in\cinf\left( S^1,{\rm Ker}\left({\rm ad\,}I_-\right)\right)\,
\right\}.
\label{5.9}\end{equation}
In terms of the  one-to-one model of ${\cal M}_c/{\cal N}$
furnished by the global gauge section  $\M_{\rm l.w.}$,
the reduced second PB is given by  the Dirac
bracket algebra of the components of $j_{\rm l.w.}$ induced from (\ref{5.4}).
This Dirac bracket algebra is  just the classical  ${\cal W}$-algebra
of \cite{bais} (see also \cite{rep}) associated to the $sl_2$ subalgebra
$\{ I_-, I_0, I_+\}\subset \G$.

A generalized KdV hierarchy of bi-Hamiltonian flows is generated on the
reduced space ${\cal M}_c/{\cal N}$ by
the commuting Hamiltonians  provided  by the
local  monodromy invariants of ${\cal L}$,
which are  determined
through the abelianization procedure
described in equations (\ref{1.9a}), (\ref{1.9b}).

\medskip

The hierarchy on $\M_{\rm red}$ resulting from the
above ``DS  type'' symmetry
reduction \cite{DS} often possesses a  residual  symmetry  that may
be used to reduce it further.
Define the subgroup  $G_R$  of ${\rm Stab}(C_-)$  by
\begin{equation}
G_R:={\rm Stab}\left(C_-\right)\cap{\rm Stab}\left(I_+\right)\cap {\rm
Stab}\label{E.1}\left(I_-\right).
\end{equation}
Let $\{ T_a\}$ denote a basis of the Lie algebra $\G_R$ of $G_R$,
\begin{equation}
\G_R={\rm Ker}\left({\rm ad\,}C_-\right)
\cap{\rm Ker}\left({\rm ad\,}I_+\right)\cap {\rm Ker
}\left({\rm ad\,} I_-\right).
\label{E.3}
\end{equation}
In fact the subgroup
$\widetilde{G_R}:=C^\infty\left(S^1,G_R\right)$ of
$\widetilde{{\rm Stab}}(C_-)$
survives the DS type symmetry  reduction.
Taking  ${\cal M}_{\rm l.w.}$ as the model of ${\cal M}_c/{\cal N}$,
the residual $\widetilde{G_R}$ symmetry acts as
\begin{equation}
(\pa + j_{\rm l.w.} + \Lambda ) \mapsto g (\pa  + j_{\rm l.w.} + \Lambda)g^{-1}
= g (  \pa + j_{\rm l.w.})g^{-1} + \Lambda,
\quad
\forall\, g\in\widetilde{G_R}.
\label{E.2}\end{equation}
These transformations  leave invariant the compatible PBs and the
commuting Hamiltonians constituting the KdV type system
on ${\cal M}_{\rm l.w.}$.
At the infinitesimal level, the $\widetilde{G_R}$ symmetry in (\ref{E.2})
is generated through the second (${\cal W}$-algebra) PB
by the components ${\rm tr}\left(T_a j_{\rm l.w.}\right)$  of
$j_{\rm l.w.}$, that is by the subset of the $sl_2$ singlet
components of $j_{\rm l.w.}$ annihilated by ${\rm ad\,} C_-$.

\medskip

The residual symmetry in (\ref{E.2}) is a continuous symmetry.
Another interesting possibility,
which is important in examples as we shall see later,
 is the presence of a discrete symmetry.
This occurs for instance in the following situation.
Let $\gamma: {\cal G} \rightarrow {\cal G}$ be an
involutive automorphism with a corresponding involution
$\Gamma: G\rightarrow G$.
In the obvious way, extend $\gamma$ to an involution of $\ell(\G)$.
Suppose now that $\Lambda$ is a grade one  regular semisimple element of
$\ell(\G)$ which is $\gamma$-invariant, $\gamma(\Lambda)=\Lambda$, and
the grading $d_{m,I_0}$ is also invariant, $\gamma(I_0)=I_0$.
Suppose furthermore that the fixed point set ${\cal G}^\gamma
\subset {\cal G}$ is a simple Lie algebra.
(All  classical Lie algebras are
fixed point sets in $gl_n$, or $sl_n$, for appropriate $\gamma$.)
The Heisenberg subalgebra
${\tilde {\cal H}}_\Lambda:= {\rm Ker}\left({\rm ad\,}\Lambda\right)$
of $\ell(\G)$ is an invariant subspace of $\gamma$, and
the fixed point set
$\tilde {\cal H}_\Lambda^\gamma\subset \tilde {\cal H}_\Lambda$ is
a Heisenberg subalgebra of $\ell(\G^\gamma)$.
We can now  perform the above DS reduction leading to  a
KdV type hierarchy
using the same element $\Lambda$ and either a system based on $\G$ or one
based on $\G^\gamma$ as the original system.

In the former case we start with the bi-Hamiltonian manifold
${\cal M}$ in (\ref{5.3}), introduce
the constrained manifold ${\cal M}_c$ in (\ref{5.7}), and end up
with ${\cal M}_{\rm red}={\cal M}_c/{\cal N}$.
The natural action of $\gamma$ on ${\cal M}$, given by
\begin{equation}
\gamma :
(\pa + J +\lambda C_-) \mapsto (\pa + \gamma(J) + \lambda C_-),
\label{E.4}\end{equation}
leaves invariant the compatible PBs on ${\cal M}$.
Since ${\cal M}_c$ is mapped to itself by $\gamma$ and also
${\cal N}$ is mapped to itself by  $\Gamma$ as $\gamma$ preserves the
grading,  the action (\ref{E.4}) induces a corresponding
action of $\gamma$  on ${\cal M}_{\rm red}$.
On account of $\gamma(I_-)=I_-$,
which may be assumed by choosing $I_-$,
the  gauge section ${\cal M}_{\rm l.w.}$ of the
${\cal N}$ orbits in ${\cal M}_c$, defined in (5.9),
is mapped to itself by  $\gamma$ in (\ref{E.4}).
Hence in terms of the model $\M_{\rm l.w.}$ of ${\cal M}_{\rm red}$
the induced action  is simply given by
\begin{equation}
\gamma: (\pa + j_{\rm l.w.} +\Lambda)
\mapsto (\pa + \gamma(j_{\rm l.w.}) + \Lambda).
\label{E.5}\end{equation}
The  action on
${\cal M}_{\rm red}={\cal M}_c/{\cal N}\simeq {\cal M}_{\rm l.w.}$
given by (\ref{E.5})
leaves invariant the compatible PBs induced from
those in (\ref{5.4}), (\ref{5.5}) by means of the DS reduction.
Recall that the Hamiltonian densities yielding the
commuting Hamiltonians of the KdV type
hierarchy on ${\cal M}_{\rm red}$ are the components of
$h(j)\in \tilde {\cal H}_\Lambda$  defining the ``abelianized''
form $(\pa + h(j)+\Lambda)$ of ${\cal L}=(\pa +j +\Lambda)\in {\cal M}_c$.
The uniqueness property of the abelianization procedure
in (\ref{1.9a}), (\ref{1.9b})
implies the equality
\begin{equation}
h\left(\gamma(j)\right)=\gamma\left(h(j)\right),
\label{E.6}\end{equation}
which  means that the Hamiltonians corresponding to the components
of $h(j)$ in ${\tilde {\cal H}}^\gamma_\Lambda$ are invariant,
and those corresponding  to
the eigenvalue $-1$  of $\gamma$ on the Heisenberg subalgebra
${\tilde {\cal H}}_\Lambda$
are ``anti-invariant'' (change sign) under the action of $\gamma$.
Since the PBs are $\gamma$-invariant,
the Hamiltonian flows on $\M_{\rm red}$ generated by the  $\gamma$-invariant
Hamiltonians   preserve the fixed point set
${\cal M}_{\rm red}^\gamma\subset {\cal M}_{\rm red}$ of $\gamma$.
(The ``anti-invariant'' Hamiltonians vanish on the
fixed point set and the
Hamiltonian flows defined by them are transversal to it.)
Therefore we can define a hierarchy on ${\cal M}^{\gamma}_{\rm red}$
by restricting the flows of  the hierarchy
generated on ${\cal M}_{\rm red}$  by  the $\gamma$-invariant Hamiltonians
to ${\cal M}^{\gamma}_{\rm red}$.
The flows of the resulting  hierarchy
are Hamiltonian with respect to the compatible PBs on the space
${\cal M}_{\rm red}^\gamma\simeq {\cal M}_{\rm l.w.}^\gamma$
obtained from those on
${\cal M}_{\rm l.w.}$ by restricting the PBs of the
$\gamma$-invariant components of $j_{\rm l.w.}$,
which may be regarded as coordinates on
${\cal M}_{\rm l.w.}^\gamma$, to this fixed point set.
We refer to the reduction  procedure just given as  ``discrete reduction''.

Using the gauge group $\N^\Gamma$ whose Lie algebra is
$\cinf\left(S^1,{\cal G}_{<0}^\gamma\right)$,
we can also perform the above discussed DS type reduction
of the system on $\M^\gamma$,
\begin{equation}
\M^\gamma= \left\{\, {\cal L}= \pa + J + \lambda C_-\, \vert\,
J\in \cinf\left(S^1,{\cal G}^\gamma\right) \,\right\}.
\label{E.7}\end{equation}
The system on ${\cal M}^\gamma$ consists  of
the compatible Poisson brackets,  defined similarly to (5.4)
and (5.5) using $\G^\gamma$ in place of $\G$,
and the monodromy invariants.
Here the  invariant scalar product ``${\rm tr}$'' on
${\cal G}^\gamma\subset \G$ is taken to be the
restriction of that  on $\G$.
Clearly, the  system on $\M^\gamma$ may be obtained by
discrete reduction from  the system on ${\cal M}$.
The discrete reduction of $\M$ to $\M^\gamma$
induces  the discrete reduction of ${\cal M}_{\rm red}$ to
$\M^\gamma_{\rm red}$.
We then have the following result.

\medskip\noindent
{\bf Proposition 5.1.}
{\em The  hierarchy on ${\cal M}_{\rm red}^\gamma$
defined as the discrete reduction
of the hierarchy on $\M_{\rm red}$ is the same as
the hierarchy  obtained from
the DS type reduction of the  system on ${\cal M}^\gamma$
using the regular semisimple element
$\Lambda\in\ell(\G^\gamma)$ and the gauge group
$\N^\Gamma=\exp\left(\cinf\left(S^1,{\cal G}_{<0}^\gamma\right)\right)$.}

\medskip\noindent
{\em Proof.}
The statement follows by an elementrary  ``diagram chasing'' argument.
{\em Q.E.D.}
\medskip

The commutativity of the diagram comprising the
two DS type reductions and the respective discrete reductions
does not depend on using the models of the DS-reduced systems
provided by the respective lowest weight gauges, since the
reduced systems have gauge independent meaning.
One
usually has other convenient gauges as well for describing  KdV
type systems
and  their ``modified'' versions.
Another possibility which is often applicable
is not to use any gauge at all for this purpose,
but rather  encode the gauge invariant information
contained in the first order
differential operator ${\cal L}\in {\cal M}_c$ in a corresponding
higher order (pseudo-)differential operator.
This will be illustrated by the examples in Subsection 5.2.
In those examples the
KdV system associated by DS reduction to a grade one  regular semisimple
element in the loop algebra of a
classical Lie algebra, realized as $\G^\gamma$ for $\G=gl_n$,
will turn out to be a discrete
reduction of a hierarchy based on $gl_n$.
In the above $\G$ was assumed to be a simple
Lie algebra, but of course the whole construction
applies to $\G=gl_n$ too.

\newpage

\subsection{Examples: Lax operators of Gelfand-Dickey type}

A traditional method for describing KdV type systems
that has proved fruitful in the past
is to find a Gelfand-Dickey type model, where the gauge
invariant dynamical variables of the system are encoded in a higher
order (pseudo-)differential Lax operator $L$.
The operator $L$ is usually derived by an ``elimination
procedure'' (see e.g.~\cite{DS,bal,FHM})
applied to the linear problem ${\cal L}\psi=0$ for
${\cal L}\in {\cal M}_c$.
The purpose of this subsection is   to derive the
Gelfand-Dickey type
pseudo-differential  Lax
operators for a subset of the
generalized KdV hierarchies resulting from the approach discussed
in Subsection 5.1.
We shall restrict ourselves to the cases for which $\G$ is a classical
Lie algebra and the regular
reductive subalgebra involved in the construction
of the Heisenberg subalgebra of $\ell(\G)$  contains only  $A$ or
$C$ type simple factors, see Table 4.
The reason for this restriction is that the elimination
procedure proves straightforwardly applicable in these cases.
The  cases involving the  subalgebras $D_{2p}$ with
the conjugacy classes $(\bar p,\bar p)\subset {\bf W}(D_{2p})$
appear  more difficult and are left aside for future work.
It will turn out that
the Lax operators obtained from the eliminitation procedure
may be also derived by suitable
restrictions from those related to $gl_n$, given in
equations (\ref{1.3}) and (\ref{1.4}).
The restriction consists in requiring the invariance of the Lax operator
under some involutive discrete symmetry.
Proposition  5.1 will be used to identify
the Poisson brackets and the commuting Hamiltonians of the
hierarchy in terms of the Gelfand-Dickey model.
We shall study in some detail the  $C_n$ and $B_n$ algebras,
and essentially give the results for  $D_n$.

\bigskip\noindent
{\bf Notations.}
Throughout this subsection,  we  use  the $2\times 2$ matrices
$\sigma$, $\tau$ defined by
\begin{equation}
\sigma:=\left(\begin{array}{cc}0&1\\-1&0\end{array}\right),
\qquad
\tau:=\left(\begin{array}{cc}1&0\\0&-1\end{array}\right),
\label{N.1}\end{equation}
and the $p\times p$ matrices $Y_p$, $\eta_p$ defined by
\begin{equation}
Y_p := {\rm diag}\left(
{p-1\over 2},{p-3\over 2},\ldots,{3-p\over 2},{1-p\over 2}\right),
\qquad
(\eta_p)_{ij}:= \delta_{i,p+1-j}\quad \forall p>1.
\label{N.2}\end{equation}
For a $p\times p$ matrix $\mu$,
$\tilde \mu := \eta_p \mu^t\eta_p$ is the transpose of $\mu$ with
respect to the antidiagonal.
As displayed  also in (1.2),
we have  the regular semisimple element $\Lambda_p\in \ell(A_{p-1})$,
\begin{equation}
\Lambda_p:=\lambda e_{p,1} +\sum_{i=1}^{p-1} e_{i,i+1}.
\label{N.3}\end{equation}
For any $p>1$ and  $s\in {\bf N}$, we fix some non-zero
$d_i\in {\bf C}$ ($i=1,\ldots,s$) satisying
$(d_i)^p\neq (d_k)^p$ for $i\neq k$ (compare with Table 1),  and
introduce the diagonal matrices
\begin{equation}
D_0 := {\rm diag}\left(d_1,\ldots,d_s\right),
\qquad
D:= {\rm diag}\left(D_0, -\tilde D_0\right),
\qquad
\Delta := -D^{-1}.
\label{N.4}\end{equation}
The $r\times r$ identity matrix is denoted by ${\bf 1}_r$
for any integer $r>1$.
Finally, the adjoint $L^\dagger$ of some matrix pseudo-differential operator
$L=\sum_{k\leq N}\alpha_k\partial^k$ is by definition
$L^{\dagger}:= \sum_{k\leq N}(-\partial)^k (\alpha_k)^t$.

\newpage

\begin{itemize}
\item {\bf Negative cycles in $C_{ps}$}
\end{itemize}

We first consider the algebra $C_{ps}$ with the conjugacy class
of ${\bf W}(C_{ps})$  associated to the signed  partition
$(\bar p,\ldots,\bar p)$.
This conjugacy class  corresponds to the regular semisimple subalgebra
$(C_p+\cdots+ C_p)\subset C_{ps}$ in Table 4.
Following the scheme outlined  in Subsection 4.1,
we first introduce the $2p\times 2p$ symplectic matrix $\Omega_{2p}$,
\begin{equation}
\Omega_{2p}:=\sigma\otimes \eta_p,\quad\hbox{that is}\quad
(\Omega_{2p})_{ij}=\epsilon(i,j) \delta_{i,2p+1-j},\quad
  \epsilon(i,j)=\left\{{\displaystyle 1
\ {\rm if}\  i<j \atop\displaystyle  -1\  {\rm if}\  i>j. }\right.
\end{equation}
The $2ps\times 2ps$ symplectic matrix $\Omega$ used to define
 $C_{ps}\subset gl_{2ps}$ is  given by
\begin{equation}
\Omega :={\bf 1}_s\otimes\Omega_{2p} =
{\rm diag}(\Omega_{2p}, \ldots ,\Omega_{2p}).
\label{symp}\end{equation}
According to (4.1) the grading operator  is $d_{2p,I_0}$
with the $2ps\times 2ps$ diagonal matrix
\begin{equation}
I_0:={\bf 1}_s\otimes Y_{2p}={\rm diag }(Y_{2p},\ldots,Y_{2p}).
\end{equation}
We also need the grade one regular semisimple element
$\Lambda^C_{2p}\in\ell(C_p)$ given by
\begin{equation}
\Lambda^C_{2p}:=\lambda e_{2p,1}+
\sum_{i=1}^{p} e_{i,i+1}- \sum_{i=p+1}^{2p-1} e_{i,i+1}.
\end{equation}
A grade one  regular semisimple  element $\Lambda\in\ell(C_{ps})$
is then furnished  by
\begin{equation}
\Lambda =D_0\otimes\Lambda^C_{2p} ={\rm diag}
\left(d_1\Lambda^C_{2p},\ldots,d_s\Lambda^C_{2p}\right).
\label{CELA}\end{equation}

Let us  perform the change of basis
that gives rise to the  permutation $P$ on the indices of
the $2ps\times 2ps$ matrices,
\begin{equation}
P(2kp+i):=2s(i-1)+k+1, \quad 1\leq i\leq 2p,\quad 0\leq k\leq s-1.
\end{equation}
This amounts to exchanging the factors in the tensor products above, i.e.,
in the new  basis the symplectic matrix is written as
$\Omega=\Omega_{2p}\otimes{\bf 1}_s$,
the grade one  regular semisimple element reads
$\Lambda=\Lambda^C_{2p}\otimes D_0$,
and the grading matrix  becomes  $I_0=Y_{2p}\otimes{\bf 1}_s$.
It will be convenient that the entries of $I_0$
are non-increasing  along the diagonal.

Now we  derive the Lax operator for the KdV system following
{}from the DS reduction. For this
we apply  the definitions of
the constrained manifold $\M_c$ in (5.7) and the gauge group
$\N$ in (5.8) to the case at hand.
We then consider the linear problem
for ${\cal L}\in {\cal M}_c$, that  is the equation
\begin{equation}
{\cal L}\psi = (\partial+j+\Lambda)\psi =0.
\label{bof}\end{equation}
Here $\psi =(\psi_1^t,\psi_2^t,\ldots,\psi_{2p}^t)^t$ is a $2ps$-vector and
the $\psi_i$ ($i=1,\ldots, 2p$) are $s$-vectors.
Equation (\ref{bof}) is covariant with respect to $\N$ if
we complement
(\ref{5.8}) with the transformation rule
\begin{equation}
e^f: \psi\mapsto e^f \psi,
\qquad \forall e^f\in \N.
\label{eff}\end{equation}
Notice that the transformation in (\ref{eff})
leaves the component $\psi_1$ invariant,
because $f$  is now given by a $2ps\times 2ps$  block-triangular
matrix  having  $s\times s$ zero blocks on and above the diagonal.
It  convenient to proceed by restricting $\L\in \M_c$
to the {\em block-diagonal gauge},
where $j$ is defined to have the form
\begin{equation}
j={\rm diag}(\theta_1,\ldots,\theta_{2p}),
\label{diag}\end{equation}
with
\begin{equation}
\theta_i\in\cinf(S^1,gl_s),
\quad
\theta_{2p+1-i}=-\theta_i^t,
\quad \forall\, i=1,\ldots, 2p.
\label{inv}
\end{equation}
Inserting $j$ in (\ref{diag}) into (\ref{bof}) yields the system
\begin{eqnarray}
&&(\partial+\theta_i)\psi_i +D_0\psi_{i+1}=0,\quad i=1, \ldots, p,
\nonumber\\
&&(\partial+\theta_i)\psi_i -D_0\psi_{i+1}=0,\quad i=p+1, \ldots, 2p-1,
\nonumber\\
&&(\partial+\theta_{2p})\psi_{2p}+\lambda D_0\psi_1 =0.
\end{eqnarray}
Upon elimination, this system leads to the eigenvalue equation
\begin{equation}
L\psi_1=\lambda\psi_1,
\label{scal}\end{equation}
where $L$ is the $s\times s$ matrix differential operator of
order $2p$ given by
\begin{equation}
L=(-1)^{p+1}D_0^{-1}(\partial+\theta_{2p})D_0^{-1}
(\partial+\theta_{2p-1})\cdots
D_0^{-1}(\partial+\theta_{1}).
\label{fac}\end{equation}
As a consequence of (\ref{inv}),  $L$ is invariant
with respect to  the operation
\begin{equation}
L\mapsto  \hat L := D_0^{-1}L^{\dagger}D_0.
\label{cond}\end{equation}
If we use an expanded form of the Lax operator $L$, we have
\begin{equation}
L=(-1)^{p+1}D_0^{-2p}\partial^{2p}+D_0^{-1}
\sum_{k=1}^{2p}(u_k\partial^{2p-k}
+\partial^{2p-k}u_k),
\label{lax}\end{equation}
where the  KdV fields  $u_k\in\cinf(S^1,gl_s)$  satisfy
$u_k^t=(-1)^k u_k$ by the  invariance property $\hat L=L$.

Since the above elimination procedure can be turned backwards,
equation (\ref{scal}) encodes all gauge invariant
information contained in the original linear problem (\ref{bof}).
It is easy to see that the KdV fields $u_k$ in (\ref{lax}) are
related by an invertible differential polynomial
substitution  to the entries of the gauge fixed current in
the lowest weight gauge of (\ref{5.9}).
The fields $\theta_i$ in (\ref{fac}) are the dynamical
variables of a ``modified'' version of the KdV hierarchy.
Expanding the factorized operator (\ref{fac})
yields a generalization  of the well-known Miura map.

The KdV system having the Lax operator $L$ in (\ref{lax}) may be interpreted
as a discrete reduction (in the sense of Subsection 5.1)
of a KdV system  based on  $gl_n$ for $n=2ps$.
In fact, the subalgebra  $C_{ps}$ of $gl_{2ps}$ is the fixed
point set of the involution $\gamma: gl_{2ps} \rightarrow gl_{2ps}$
defined by
\begin{equation}
\gamma: X\mapsto \gamma(X):= -\Omega^{-1} X^t \Omega
\qquad \forall\, X\in gl_{2ps},
\label{gamma}\end{equation}
and the element
$\Lambda \in \ell(C_{ps})\subset \ell(gl_{2ps})$ given in (\ref{CELA})
is  also a grade one regular semisimple element of $\ell(gl_{2ps})$
(and of $\ell(A_{2ps-1})$).
{}From this point of view $\Lambda$ is associated to the partition
$(2p,\ldots,2p)$   of $n=2ps$ representing a regular conjugacy
class in ${\bf W}(A_{2ps-1})$.
Performing the DS reduction using $gl_{2ps}$ instead of $C_{ps}$
leads to a KdV system whose  Lax operator has the form in (\ref{lax}), but
 with {\em arbitrary} $u_k\in\cinf(S^1,gl_s)$.
The related modified KdV system is given by the
operator ({\ref{fac}) with unrestricted $\theta_i\in \cinf(S^1,gl_s)$.
Proposition 5.1 and what is known about
the $gl_n$ case \cite{FHM} enables  us to
give a more detailed description of the present
generalized KdV hierarchy in the Gelfand-Dickey framework.
We next explain this in detail.

Let $M$ be the manifold  of Lax operators $L$ of the form
in (\ref{lax}) with arbitrary KdV fields $u_k\in\cinf(S^1,gl_s)$.
Recall from \cite{FHM} that the compatible PBs
 on  $M$, regarded  as a model of
the DS-reduced space ${\cal M}_{\rm red}$ associated to $gl_{2ps}$,
are the standard
first and second matrix Gelfand-Dickey PBs \cite{GD,Di,Ad}
defined respectively by
\begin{equation}
\{ f_A, f_B\}^{(1)}(L)={\rm Tr}\left( L
\left([A_+,B_+]-[A_-,B_-]\right)\right),
\label{gd1}\end{equation}
\begin{equation}
\{ f_A, f_B\}^{(2)}(L)={\rm Tr}\left(BL(AL)_+-B(LA)_+L\right).
\label{gd2}\end{equation}
Here ${\rm Tr}$ is the  Adler trace \cite{Ad} of matrix
pseudo-differential operators (PDOs)  given by
\begin{equation}
{\rm Tr}( A):= \int_{S^1} {\rm tr\,} {\rm res} \left(A\right),
\quad
{\rm res}\left(A\right):= A_{-1}
\qquad
\forall\, A=\sum_{k\leq k_0} A_k \pa^{k},
\quad
A_k\in\cinf( S^1,gl_s).
\label{Tr}\end{equation}
For an arbitrary PDO $A$,  we use
the splitting $A=A_++A_-$ into
parts containing non-negative and negative powers of $\pa$, respectively.
In formulas (\ref{gd1}), (\ref{gd2}) $f_A$ is the linear function on $M$
defined  by $f_A(L):= {\rm Tr}\left(AL\right)$ for any fixed
$s\times s$ matrix PDO $A$.

We have the  discrete symmetry given by  the Poisson mapping
\begin{equation}
\hat\gamma: M\rightarrow M, \qquad
\hat\gamma(L):=\hat L=D_0^{-1}L^\dagger D_0,\quad  \forall L\in M.
\label{condd}\end{equation}
The symmetry  $\hat \gamma$ is induced from the action (\ref{E.4})
of $\gamma$ in (\ref{gamma}) on the constrained manifold
of the DS reduction considered for $gl_{2ps}$.
This is easily seen with the aid of the corresponding  block-diagonal gauge,
whose  gauge section  is  mapped to itself by $\gamma$.
The phase space of the ``discrete reduced''
hierarchy is the fixed point set $M^{\hat\gamma}\subset M$ of
$\hat \gamma$.
Proposition 5.1  implies that
the induced PBs on the fixed point set $M^{\hat\gamma}$,
which is  a model of ${\cal M}_{\rm red}^\gamma$,
are given by  formulas (\ref{gd1}) and (\ref{gd2}),
where $A$ and $B$ have to be restricted
to  PDOs that are  anti-symmetric with respect to
the transformation $\hat \gamma$.
Indeed, if $\hat\gamma(A):= D_0^{-1} A^\dagger D_0 =-A$,
then $f_A(\hat\gamma(L))=f_A(L)$.

The commuting Hamiltonians of the hierarchy on $M$
induced by the  DS reduction may be  obtained as follows \cite{FHM}.
First one  has to diagonalize $L\in M$ in the algebra
of PDOs, i.e., for any $L$ one has to determine
a diagonal PDO $L_d$,
\begin{equation}
L_d=(-1)^{p+1} D_0^{-2p} \pa^{2p} +\sum_{k=1}^\infty a_k \pa^{2p-k},
\quad\hbox{with}\quad  a_k\quad\hbox{diagonal matrix}\quad \forall k,
\label{Ld}\end{equation}
for which
\begin{equation}
L=g L_d g^{-1},\qquad
g={\bf 1}_s +\sum_{k=1}^{\infty} g_k \pa^{-k},
\quad\hbox{with}\quad  g_k\quad\hbox{off-diagonal matrix}\quad \forall k.
\label{diago}\end{equation}
By (\ref{Ld}), (\ref{diago}), $L_d(L)$ and $g(L)$
are uniquely determined (differential polynomial) functions of $L\in M$.
The commuting Hamiltonians  are  then provided by
\begin{equation}
H_{0,i}(L):= \int_{S^1} \left(u_1\right)_{ii} ,
\qquad
\forall\,i=1,\ldots,s,
\label{ham0}\end{equation}
\begin{equation}
H_{k,i}(L):= \int_{S^1}{\rm res\,} \left(L_d(L)\right)_{ii}^{k/{2p}},
\qquad
\forall\,i=1,\ldots, s,\quad k=1,2,\dots\,,
\label{hamk}\end{equation}
where $\left(L_d(L)\right)^{1/{2p}}$ is a fixed
$2p$th root of $L_d(L)$.
Thanks to the  uniqueness property of the diagonalization procedure
in (\ref{Ld}), (\ref{diago}) and the identity
${\rm Tr}\left( A^\dagger\right)=-{\rm Tr}\left(A\right)$,
we can verify
\begin{equation}
H_{k,i}\left({\hat\gamma}\left(L\right)\right)=(-1)^{k+1} H_{k,i}(L),
\qquad
\forall\,  i=1,\ldots, s, \quad k=0,1,\ldots\,.
\label{hsym}\end{equation}

According to Proposition 5.1,
the commuting Hamiltonians of the discrete reduced
hierarchy on $M^{{\hat\gamma}}\simeq \M_{\rm red}^\gamma$
are furnished by  the restrictions of the
$\hat \gamma$-invariant Hamiltonians on
$M\simeq {\cal M}_{\rm red}$.
We see from (\ref{hsym}) that the invariant Hamiltonians are now
the $H_{k,i}(L)$ for $k$ any {\em odd} natural number.
This completes our description of the PDO model
of the generalized KdV  hierarchy following from DS reduction
in the case $(\bar p,\ldots,\bar p)\subset {\bf W}(C_{ps})$.
The result is analogous to  the $s=1$  ``scalar case'',
for which the $C_p$-type DS hierarchy is the self-adjoint
reduction of the  $gl_{2p}$-type Gelfand-Dickey
($n$-KdV for $n=2p$) hierarchy \cite{DS}.

\begin{itemize}
\item {\bf Positive cycles in $C_{ps}$}
\end{itemize}

We now turn to the case of positive cycles of odd length,
$(p,\ldots,p)$ with $p=2q+1$, in $C_{ps}$.
The regular semisimple subalgebra associated in Table 4 to this
conjugacy class of ${\bf W}(C_{ps})$  is
$(A_{p-1}+\cdots + A_{p-1})\subset C_{ps}$
The symplectic matrix $\Omega$ is still given by  (\ref{symp}).
The grading of $\ell(C_{ps})$  is now defined by
the operator $d_{p,I_0}$  with $I_0:={\bf 1}_{2s}\otimes Y_p$.
Using  (\ref{N.1})--(\ref{N.4}),
the grade one regular semisimple element $\Lambda\in\ell(C_{ps})$ is
given as $\Lambda=D_0\otimes\tau\otimes\Lambda_p$.

Let us perform the permutation
\begin{equation}
{\makebox[2.5in][l]{$P(2kp+i):=2s(i-1)+k+1,$} \atop
\makebox[2.5in][l]{$P(2kp+p+i):=2si-k,$}}
 1\leq i\leq p,\quad 0\leq k\leq s-1.
\label{perm2}\end{equation}
After this permutation, the symplectic matrix
writes as  $\Omega=\eta_p\otimes \Omega_{2s}$ and
the grading matrix  becomes $I_0=Y_p\otimes {\bf 1}_{2s}$,
which has non-increasing entries along the diagonal.
Finally, with $D$ given in (\ref{N.4}), we have
\begin{equation}
\Lambda=\Lambda_p\otimes D.
\label{Lamplus}\end{equation}

Like in the previous case, we consider the linear problem
(\ref{bof}).
Now the $2ps$-vector $\psi$ is decomposed as
$\psi =(\psi^t_1,\ldots,\psi^t_p)^t$ in terms of
the $2s$-vectors $\psi_i$ for $i=1,\ldots,p$.
In the block-diagonal gauge  $j$ has the form
\begin{equation}
j={\rm diag}(\theta_1,\ldots,\theta_p),
\label{plusdia}\end{equation}
with
\begin{equation}
\theta_i\in\cinf\left(S^1,  gl_{2s}\right),
\qquad  \theta_i =-\Omega_{2s}\theta_{p+1-i}^t\Omega_{2s}^{-1}
\qquad \forall\, i=1,\ldots, p.
\label{condlus}\end{equation}
Combining  (\ref{bof}) with (\ref{Lamplus}), (\ref{plusdia}),
we obtain the system
\begin{eqnarray}
&& (\partial+\theta_i)\psi_i+D\psi_{i+1}=0,
\quad 1\leq i\leq p-1,\nonumber\\
&& (\partial+\theta_p)\psi_p+\lambda D\psi_{1}=0.
\end{eqnarray}
By elimination, we then get  the eigenvalue equation
$L\psi_1=\lambda\psi_1$, where
the $2s\times 2s$ matrix Lax operator $L$ is given by
\begin{equation}
L=\Delta(\partial+\theta_p)\cdots\Delta(\partial+\theta_1),
\label{fac1}\end{equation}
 with  $\Delta$ defined in  (\ref{N.4}).
On account of (\ref{condlus})
and $\Omega_{2s} \Delta^t\Omega_{2s}^{-1}=-\Delta$ ,
$L$ in (\ref{fac1}) is invariant  with respect to the transformation
\begin{equation}
L\mapsto \hat L:=
\Delta\Omega_{2s}L^{\dagger}\Omega_{2s}^{-1}\Delta^{-1}.
\label{cond1}\end{equation}
If we write  the Lax operator in expanded form as
\begin{equation}
L=\Delta^p\partial^p+\Delta\sum_{k=1}^p(u_k\partial^{p-k}+
\partial^{p-k}u_k),\label{exp1}\end{equation}
then the invariance property  $L=\hat L$ yields
$u_k=(-1)^{k}\Omega_{2s} u^t_k\Omega_{2s}^{-1}$.

Similarly to  the previous example, we see that the
KdV system possessing the Lax operator in (\ref{exp1}) is a discrete
reduction of a system of the type in (\ref{1.3}),
which is based on $gl_n$ with the partition $(p,\ldots,p)$ of $n=2ps$.
It follows that the compatible PBs
of the KdV system obtained from the  DS reduction
 are given by  (\ref{gd1}),  (\ref{gd2}),
where $A$ and $B$  have to be restricted
to PDOs that  are  anti-symmetric with respect to the
discrete symmetry in (\ref{cond1}),
\begin{equation}
\Delta\Omega_{2s}A^{\dagger}\Omega_{2s}^{-1}\Delta^{-1}=-A,
\qquad
\Delta\Omega_{2s}B^{\dagger}\Omega_{2s}^{-1}\Delta^{-1}=-B.
\end{equation}
Before the discrete reduction, i.e., on the space of Lax operators
of the form in  (\ref{exp1}) but with
arbitrary  coefficients  $u_k\in \cinf(S^1,gl_{2ps})$,
the commuting Hamiltonians are
$H_{0,i}(L)$  defined like in (\ref{ham0})
and $H_{k,i}(L)$  defined by
\begin{equation}
H_{k,i}(L):= \int_{S^1}{\rm res\,} \left(L_d(L)\right)_{ii}^{k/{p}},
\qquad
\forall\,i=1,\ldots, 2s,\quad k=1,2,\dots\,.
\label{newhamk}\end{equation}
Here $\left(L_d(L)\right)^{1/{p}}$ is a fixed
$p$th root of the diagonal PDO $L_d(L)$ determined analogously
to (\ref{diago}).
Choosing the leading term of $\left(L_d(L)\right)^{1/{p}}$ to be
$\Delta \pa$,  we  find the transformation property
\begin{equation}
H_{k,i}(\hat L)=-H_{k,2s+1-i}(L),
\qquad
\forall\, i=1,\ldots, 2s,\quad k=0,1,\dots\,.
\label{hamtraf}\end{equation}
Therefore the  Hamiltonians of the KdV system based on
$gl_{2ps}$ that are invariant with respect to the
discrete symmetry in (\ref{cond1}) are furnished by
\begin{equation}
H^+_{k,i}(L):= H_{k,i}(L)-H_{k,2s+1-i}(L),
\qquad
\forall\, i=1,\ldots, s,\quad k=0,1,\ldots\,.
\label{haminv}\end{equation}
As a consequence of Proposition 5.1,
the Hamiltonians  obtained by inserting the Lax operator $L$ in (\ref{exp1})
into (\ref{haminv})  coincide with those
resulting  from  ``abelianization''  in the DS reduction realization
of the generalized  KdV system associated to
$(p,\ldots,p)\subset  {\bf W}(C_{ps})$.

\begin{itemize}
\item {\bf Positive cycles in $D_{ps}$}
\end{itemize}

The case of positive cycles  of odd length,
$(p,\ldots,p)$ with  $p=2q+1$,
in ${\bf W}(D_{ps})$ is very similar.
We end up with a Lax  operator $L$ that has the
factorized form in (\ref{fac1}), where the matrices $\theta_i$ now
satisfy $\theta_i=-\tilde\theta_{p+1-i}$.
Thus the invariance property of $L$ is
\begin{equation}
\hat L=L \qquad \hbox{for}\qquad
L\mapsto \hat L:=\Delta\eta_{2s}L^{\dagger}\eta_{2s}\Delta^{-1}.
\label{lastcond}\end{equation}
The expanded form of the Lax  operator can be written
as in (\ref{exp1}), where the $2s\times 2s$ matrix KdV fields
 $u_k$ are  now subject to  $u_k=(-1)^{k}\tilde u_k$.
This KdV system is another discrete reduction of the
system  based on  $gl_n$ with the partition
$(p,\ldots, p)$ of $n=2ps$.
The PBs of this system following from the DS reduction
can be  obtained from the
Gelfand-Dickey PBs in (\ref{gd1}), (\ref{gd2})
by restricting $A$ and $B$ to be anti-symmetric PDOs with respect
to the transformation in (\ref{lastcond}).
The commuting Hamiltonians
can be characterized analogously to the preceding example.

\begin{itemize}
\item {\bf Positive cycles in $B_{ps}$}
\end{itemize}

Now we deal with the case of positive cycles of odd length
$(p,\ldots,p)\subset {\bf W}(B_{ps})$, $p=2q+1$.
The corresponding  regular semisimple subalgebra is given by
$(A_{p-1}+\cdots + A_{p-1})\subset B_{ps}$.
The $(2ps+1)\times (2ps+1)$ matrix $\eta$
defining the $B_{ps}$-invariant symmetric form can  be taken to be
\begin{equation}
\eta:=\left(\begin{array}{cc}{\bf 1}_s\otimes\eta_{2p}&0\\
0&1\end{array}\right).\end{equation}
The grading of $\ell(B_{ps})$ is defined  by the operator $d_{p,I_0}$ with
\begin{equation}
I_0:=\left(\begin{array}{cc}{\bf 1}_{2s}\otimes Y_p & 0\\
0 & 0\end{array}\right).
\end{equation}
The relevant  grade one regular semisimple element
$\Lambda\in \ell(B_{ps})$ can be written as
\begin{equation}
\Lambda =\left(\begin{array}{cc}D_0\otimes\tau\otimes\Lambda_p &0\\
0 & 0\end{array}\right).
\end{equation}
For the notations, see (\ref{N.1})--(\ref{N.4}).

Let us change the basis using $P$ in (\ref{perm2}) to permute
the first $2ps$ indices together  the prescription
$P(2ps+1) := 2ps+1$ for the last index.
The matrix of the symmetric form left invariant by
$B_{ps}\subset gl_{2ps+1}$ then becomes
\begin{equation}
\eta=\left(\begin{array}{cc}\eta_{2ps} &0\\
0&1\end{array}\right).\end{equation}
The grading matrix reads
\begin{equation}
I_0=\left(\begin{array}{cc}Y_p\otimes {\bf 1}_{2s} & 0\\
0 & 0\end{array}\right).
\label{I0}\end{equation}
The grade one regular  element takes the form
\begin{equation}
\Lambda =\left(\begin{array}{cc}\Lambda_p\otimes D&0\\
0 & 0\end{array}\right).
\label{Bamba}\end{equation}

In the linear problem (\ref{bof}) the vector $\psi$
may be now decomposed as
$\psi=(\psi_1^t,\ldots,\psi_p^t,\phi)^t$,
where the $\psi_i$ ($i=1,\ldots,p$) are $2s$-vectors and $\phi$ is the last
component of $\psi$.
We now define the ``block-diagonal'' gauge by restricting
the $(2s+1)\times (2s+1)$ matrix valued field
$j\in\cinf(S^1,B_{ps})$ in  $\L=(\pa + j+ \Lambda)\in \M_c$
to have  the form
\begin{equation}j=\left(\begin{array}{cccccc} \theta_1 &&&&&\\
&\ddots&&&&\\ &&\theta_{q+1}&&& b\\ &&&\ddots &&\\&&&&\theta_p &\\
&&c^t&&&0\end{array}\right).
\label{block}\end{equation}
The non-vanishing entries of $j$ in (\ref{block})
have grade zero with respect to $I_0$ in (\ref{I0}) and satisfy
\begin{equation}
\theta_i\in\cinf(gl_{2s},S^1),\quad
\theta_i=-\tilde\theta_{p+1-i}\quad
\forall i=1,\ldots, p,
\qquad
b, c \in \cinf(S^1, {\bf C}^{2s}),\quad
c=-\eta_{2s}b.
\label{tricky}\end{equation}
Substituting  (\ref{Bamba}), (\ref{block}) into ({\ref{bof}),
we obtain the system
\begin{eqnarray}&&
(\partial+\theta_i)\psi_i+D\psi_{i+1}=0,\quad i=1,\ldots,q,\,q+2,\ldots,2q,
\nonumber\\&&
(\partial+\theta_{q+1})\psi_{q+1}+D\psi_{q+2}+b\phi=0,
\nonumber\\&&
(\partial+\theta_p)\psi_p+\lambda D\psi_{1}=0,
\nonumber\\&&
\partial\phi+c^t\psi_{q+1}=0.
\label{system}\end{eqnarray}
The component $\phi$ may be eliminated using the last equation, which yields
\begin{equation}
\phi=-\partial^{-1}c^t\psi_{q+1}.
\label{eli}\end{equation}
Plugging (\ref{eli}) back into (\ref{system}),
further elimination leads to the eigenvalue equation
\begin{equation}
L\psi_1=\lambda\psi_1,
\end{equation}
where $L$ is the
following $2s\times 2s$ matrix {\em pseudo-differential} operator:
\begin{equation}
L=\Delta(\partial+\theta_p)\cdots\Delta(\partial+\theta_{q+2})\Delta
\left[\partial+\theta_{q+1}-b\partial^{-1}c^t\right]
\Delta(\partial+\theta_q)\cdots
\Delta(\partial+\theta_1).
\label{laxtheta}\end{equation}
Because of (\ref{tricky}),
$L$ in (\ref{laxtheta})  has the invariance property
\begin{equation}
\hat L=L
\qquad\mbox{for}\qquad
L\mapsto\hat L:= \Delta\eta_{2s}L^{\dagger}\eta_{2s}\Delta^{-1},
\qquad (\Delta=-D^{-1}).
\label{plus}\end{equation}
The Lax  operator given by (\ref{laxtheta}) can be written
in expanded form as
\begin{equation}
L=\Delta^p\partial^p+\Delta\sum_{k=1}^p(u_k\partial^{p-k}+\partial^{p-k}u_k)
-\Delta z_+\partial^{-1}z_-^t,
\label{LAXA}\end{equation}
\begin{equation}
u_k\in\cinf(S^1,gl_{2s}),\quad  u_k=(-1)^k\tilde u_k
\quad \forall k=1,\ldots,p,
\quad  z_+, z_-\in \cinf(S^1, {\bf C}^{2s}),
\quad  z_-=-\eta_{2s}z_+.
\label{LAXB}\end{equation}

The  above Lax operator
can be also derived  by performing the elimination
on the linear problem (\ref{bof}) in a DS gauge.
For this  it is convenient to consider the
gauge section ${\cal M}_{\rm DS}\subset {\cal M}_c$
which by definition   consists of the first order
differential operators  ${\cal L}=(\pa + j_{\rm DS} + \Lambda)$ with
\begin{equation}
j_{\rm DS}:=\left(\begin{array}{cccccc}
v_1&&&&&\\
v_2&&&&&\\
\vdots&&&&&\\
v_{p-1}&&&&&\\
v_p& -\tilde v_{p-1}&\cdots& -\tilde v_2 & -\tilde v_1 & z_+\\
z_-^t&{}&{}&{}&{}&{}\end{array}\right),
\label{furi}\end{equation}
where  $v_k\in\cinf(S^1,gl_{2s})$  subject to $v_k=(-1)^k \tilde v_k$,
and $z_\pm$ are given in (\ref{LAXB}).
The gauge section $\M_{\rm DS}$ is  a one-to-one model of
the reduced  space ${\cal M}_{\rm red}={\cal M}_c/{\cal N}$ following
{}from the DS reduction in the  present case.
The fields $v_k$ in (\ref{furi})
and the $u_k$ in (\ref{LAXA})  are related by an invertible differential
polynomial substitution, but the field $z_-$ appears only
in quadratic combinations in the
expression (\ref{LAXA}) of $L$.
This means that the manifold  of Lax operators $L$ in (\ref{LAXA})
is now {\em not} a one-to-one model of the  space $\M_{\rm red}$.
A  convenient parametrization of  ${\cal M}_{\rm red}$
is furnished by the set of all pairs $(L_+,z_-)$, where $L_+$
is the differential operator part of $L$ in (\ref{LAXA})
and $z_- \in\cinf(S^1, {\bf C}^{2s})$.
This is somewhat similar  to the situation found in \cite{DS}
for  the principal case of the $D_n$ algebras, for which the
Lax oparators are  skew-symmetric scalar pseudo-differential operators
having a negative part of the form $z \pa^{-1} z$ with
$z\in\cinf\left(S^1, {\bf C}\right)$.

Finally,  we note that the above KdV system  associated by DS reduction
to the conjugacy class $(p,\ldots,p)\subset {\bf W}(B_{ps})$ can be
viewed as a discrete reduction of a KdV system
based on $gl_{2ps+1}$ with the corresponding partition $(p,\ldots,p,1)$,
where $p=2q+1$ occurs $2s$ times.
The phase space of the  system based on $gl_{2ps+1}$ consists
of the quadruples $(L_+, y_+, y_-, w)$  appearing  in (\ref{1.4}).
The  PBs and the commuting Hamiltonians are
described in these variables in \cite{FM}.

\begin{itemize}
\item {\bf Positive cycles plus a 1-cycle in $D_{ps+1}$}
\end{itemize}

The case of positive cycles of odd length $p=2q+1$ plus a 1-cycle
$(p,\ldots,p,1)\subset {\bf W}(D_{ps+1})$ resembles the last one.
Without entering into details, let us give the form of the
$(2ps+2)\times (2ps+2)$  matrix valued field $j$ in the
``block-diagonal'' gauge,
\begin{equation}j=\left(\begin{array}{cccccc} \theta_1 &&&&&\\
&\ddots&&&&\\ &&\theta_{q+1}&&& b\\ &&&\ddots &&\\&&&&\theta_p &\\
&&c^t&&&d\end{array}\right).\end{equation}
Here the $\theta_i$ ($i=1,\ldots, p$) are $2s\times 2s$ matrices satisfying
$\theta_i=-\tilde\theta_{p+1-i}$, $b$ and $c$ are rectangular $2s\times 2$
matrices related by $c=-\eta_{2s}b\eta_2$, and $d$ is a $2\times 2$ matrix
constrained by $d=-\tilde d$.
The corresponding  pseudo-differential
Lax operator $L$ is given in factorized form as
\begin{equation}
L=\Delta(\partial+\theta_p)\cdots\Delta (\partial+\theta_{q+2})\Delta
\left[ \partial+\theta_{q+1}-b(\partial+d)^{-1}c^t\right]
\Delta(\partial+\theta_q)\cdots\Delta(\partial+\theta_1),
\label{lastfac}\end{equation}
with $\Delta$ in (\ref{N.4}).
The operator $L$ in (\ref{lastfac}) enjoys the invariance
property (\ref{plus}) and  can be expanded as
\begin{equation}
 L=\Delta^p\partial^p+\Delta\sum_{k=1}^p(u_k\partial^{p-k}+\partial^{p-k}u_k)
-\Delta z_+(\partial+d)^{-1}z_-^t,
\end{equation}
where the $2s\times 2s$ matrices $u_k$ satisfy $u_k=(-1)^k\tilde u_k$
and the rectangular $2s\times 2$ matrices $z_+$ and $z_-$ are related by
$z_-=-\eta_{2s}z_+\eta_2$.

The generalized KdV system at hand  is  related to a system
based on $\G:=gl_n$ with the partition $(p,\ldots,p,1,1)$  of $n=2ps+2$
by means of  an involution  $\gamma:\G\rightarrow \G$ for which
$\G^\gamma=D_{ps+1}$.
If there are more than one extra 1-cycles contained
in the partition $n$, then graded
{\em regular} semisimple elements do not exist in the corresponding
Heisenberg subalgebra of $\ell(gl_n)$.
However, in the cases $(p,\ldots,p,1,\ldots,1)$ --- with
an arbitrary  number of 1-cycles ---
the DS reduction
still goes through without any
difficulty using  a grade one semisimple element  from  the
Heisenberg subalgebra.
The resulting KdV type hierarchies are studied in  \cite{FM}.

\newpage

\section{Some remarks on non-abelian Toda systems}
\setcounter{equation}{0}

In the preceding section we associated   generalized KdV
systems to  grade one regular semisimple
elements of $\ell(\G)$.
For completeness, below we wish to present the
well-known  definition  of corresponding
 ``non-abelian affine Toda'' systems,
and  work out an example.

To obtain a non-abelian\footnote{The {\em abelian}
affine Toda model is related to the principal Heisenberg subalgebra
as is well-known.}
affine Toda model,  consider a grade $1$ and a
grade $-1$ regular semisimple element, $\Lambda$ and $\bar \Lambda$, from
some non-principal
Heisenberg subalgebra of $\ell(\G)$.
The grading is  given by the operator $d_{m,I_0}$ in (4.1).
For simplicity we here assume that
${\rm ad} I_0$ has only {\em integral} eigenvalues.
Similarly to equations (5.1), (5.2),
for $\Lambda$ and $\bar \Lambda$  given by
\begin{equation}
\Lambda = I_+ + \lambda C_-,
\qquad
\bar \Lambda = \bar I_- + \lambda^{-1} \bar C_+,
\label{5.10}\end{equation}
we  suppose that
\begin{equation}
[C_-, \G_{<0}]=\{ 0\},
\qquad
[\bar C_+, \G_{>0}]=\{ 0\}.
\label{5.11}\end{equation}
The non-abelian affine Toda equation is a relativistically invariant
field equation for a field $g(x,t)$ that varies in a connected
(non-abelian)  Lie group
$G_0$ generated by the grade zero Lie subalgebra $\G_0\subset \G$.
It is postulated to be the zero curvature equation
\begin{equation}
[{\cal L}_+, {\cal L}_-]=0,
\label{5.12}\end{equation}
with
\begin{equation}
{\cal L}_+:=\pa_++g^{-1}\pa_+ g + \Lambda,
\qquad
{\cal L}_-:=\pa_- + g^{-1}\bar \Lambda g,
\label{5.13}\end{equation}
where $\pa_\pm:=(\pa_x \pm \pa_t)$.
More explicitly,
the field equation (\ref{5.12}) reads
\begin{equation}
\pa_-(g^{-1}\pa_+ g)=[I_+, g^{-1}\bar I_- g] +[C_-, g^{-1} \bar C_+ g].
\label{5.14}\end{equation}
This is a deformation of the non-abelian conformal Toda equation
obtained from (\ref{5.14}) by omitting the second  term on the
right hand side.
The model admits two infinite series of  conserved local currents,
which may be obtained with the aid of the abelianization of
${\cal L}_x = {\cal L}_+ - {\cal L}_-$ and that of
${\tilde {\cal L}}_x:={\tilde{\cal  L}}_+ - {\tilde {\cal L}}_-$,
respectively, where the operators
\begin{equation}
\tilde {\cal L}_+:=\pa_++ g \Lambda g^{-1},
\qquad
\tilde {\cal L}_-:=\pa_- -\pa_-g g^{-1}+ \bar \Lambda
\label{5.15}\end{equation}
enter the alternative zero curvature representation
\begin{equation}
[\tilde {\cal L}_+, \tilde {\cal L}_-]=0
\label{5.16}\end{equation}
of the field equation (\ref{5.14}).

The models defined by (6.1)--(6.4)  are special cases of
those  proposed by Leznov and Saveliev in \cite{LS}.
They  are distinguished by the applicability of the
abelianization procedure described in  (1.9),(1.10).
It is well-known \cite{DS,Wi,Oli,under,McI,Prin1}
 that infinitely many conserved  local currents
exist also in the non-abelian affine Toda models associated
to grade $\pm 1$ semisimple, not necessarily regular
elements from $\ell(\G)$.
In general the conserved local currents are labelled
by the  basis elements of the {\em centre of the centralizer} of
$\Lambda$ ($\bar \Lambda$) with non-positive (non-negative) grades.

Suppose that we consider a regular conjugacy class
of the Weyl group that has the  product structure in (4.4).
The corresponding  Toda model then will have the interpretation
as a ``coupled system'' containing  the Toda systems
associated to grade $\pm 1$ regular elements from the primitive
Heisenberg subalgebras $\tilde {\cal H}_{k, \hat w_k}\subset \ell(\G_k)$
for $k=1,\ldots, r$
(see (4.4)--(4.10)), which  are coupled together by means of
certain extra fields.
The extra fields correspond to the part of $\G_0$  outside the
regular semisimple subalgebra given in (4.5).
It is easy to see that
the extra fields can be
consistently set to zero in the field equation (6.5),
which then reduces to a decoupled set of Toda equations associated
to the primitive conjugacy classes $[w_k]\subset {\bf W}(\G_k)$.

We now wish to elaborate the non-abelian affine
Toda equation
(\ref{5.14}) for the two negative cycles case $(\bar p,\bar p)$ in
$D_{2p}$ for any $p\geq 2$.
The motivation for considering this series of  examples is that for
the classical Lie algebras
$(\bar p, \bar p)\subset {\bf W}(D_{2p})$
are  the only conjugacy
classes of the Weyl group which are regular, primitive and  different from
a Coxeter class.
Choosing all constants $a_i, b_i$ in (\ref{4.27}) to be $1$ for simplicity,
the grade $1$ generators of the corresponding
Heisenberg subalgebra are
\begin{eqnarray}
&&\Lambda_{1,1}=\lambda(e_{2p,1}-e_{2p+1,2})
+\sum_{k=1}^p e_{k,k+1} - \sum_{k=1}^p e_{p+k,p+k+1},
\label{T1}\\
&&\Lambda_{1,2}=
\lambda(e_{4p,1}-e_{2p+1,2p+2}) +(e_{4p,2p+1}-e_{1,2p+2})
+\sum_{k=1}^{p-1} e_{2p+1+k,2p+2+k}
-\sum_{k=1}^{p-1} e_{3p+k,3p+k+1},
\nonumber\end{eqnarray}
and the  grade $-1$ generators,
$\Lambda_{-1,i}\sim \lambda^{-1} (\Lambda_{1,i})^{2p-1}$, are
\begin{equation}
\Lambda_{-1,1}=\lambda^{-1}(e_{1,2p} -e_{2,2p+1}) +(e_{2,1}-e_{2p+1,2p})
+2\sum_{k=1}^{p-1} e_{2+k,1+k} -2\sum_{k=1}^{p-1} e_{p+1+k,p+k},
\label{T2}\end{equation}
$$
\Lambda_{-1,2}=\lambda^{-1}(e_{1,4p} -e_{2p+2,2p+1}) +
(e_{2p+1,4p}-e_{2p+2,1})
+2\sum_{k=1}^{p-1} e_{2p+2+k,2p+1+k} -
2\sum_{k=1}^{p-1} e_{3p+1+k,3p+k}.
$$
These formulas are valid in the basis where the symmetric form
 $\eta$ and the grading $K$ are given by  (\ref{4.20}) and (\ref{4.21}),
and it is convenient to permute the basis so that in the new basis
they take the following block-form:
\begin{eqnarray}
&{}&K={\rm diag}\left(p, (p-1) {\bf 1}_2,\ldots,
-(p-1) {\bf 1}_2, -p\right),
\label{T.3}\\
&{}&\eta = {\rm antidiag}\left(1, {\bf 1}_2,  \ldots, {\bf 1}_2, 1\right).
\nonumber\end{eqnarray}
According to the grading defined by  $K$,  we can write all  matrices in a
$(2p+1)\times (2p+1)$ block-form, with the various blocks being
$2\times 2$ matrices and $2$-component column or row vectors, respectively.
In  order to write down the grade $\pm 1$ regular elements
$\Lambda=d_1 \Lambda_{1,1} + d_2 \Lambda_{2,1}$ and
$\bar \Lambda :=\bar d_1 \Lambda_{-1,1} +\bar d_2 \Lambda_{-1,2}$,
it is useful to introduce
\begin{equation}
\alpha:=\left(\matrix{d_1\cr d_2\cr}\right),
\quad
\beta:=\left(\matrix{d_1\cr-d_2\cr}\right),
\quad
D_0:=\left(\matrix{d_1&0\cr 0&d_2\cr}\right),
\label{T.4}\end{equation}
and
\begin{equation}
\bar \alpha:=\left(\matrix{\bar d_1\cr \bar d_2\cr}\right),
\quad
\bar\beta:=\left(\matrix{\bar d_1\cr-\bar d_2\cr}\right),
\quad
\bar D_0:=2\left(\matrix{\bar d_1&0\cr 0&\bar d_2\cr}\right).
\label{T5}\end{equation}
Using this notation,  in the new basis we have
\begin{eqnarray}
&{}&\Lambda = e_{1,2}\otimes \beta^t +\sum_{k=2}^{p} e_{k,k+1} \otimes D_0
-\sum_{k=p+1}^{2p} e_{k,k+1} \otimes D_0 -e_{2p,2p+1}\otimes \beta
\nonumber\\
&{}&\phantom{\Lambda =}
+\lambda\left( e_{2p,1}\otimes \alpha -e_{2p+1,2}\otimes \alpha^t\right),
\label{T6}\end{eqnarray}
and
\begin{eqnarray}
&{}&\bar \Lambda = e_{2,1}\otimes \bar \beta +\sum_{k=2}^{p} e_{k+1,k}
\otimes \bar D_0
-\sum_{k=p+1}^{2p} e_{k+1,k} \otimes \bar D_0 -
e_{2p+1,2p}\otimes \bar\beta^t
\nonumber\\
&{}&\phantom{\bar\Lambda =}
+\lambda^{-1}\left( e_{1,2p}\otimes \bar\alpha^t -e_{2,2p+1}\otimes
\bar\alpha\right).
\label{T7}\end{eqnarray}
We write the group element $g\in G_0$ in the block-diagonal form
\begin{equation}
g=\sum_{k=1}^{2p+1} e_{k,k} \otimes g_k,
\label{T8}\end{equation}
where  $g_1, g_{2p+1}\in GL(1)$ and $g_k\in GL(2)$ otherwise,
with the condition $g^t \eta g=\eta$ translating into
\begin{equation}
g_{2p+2-l}=(g_l^{-1})^t
\quad\hbox{for}\quad l=1,\ldots, p+1.
\label{T9}\end{equation}
Then the
 non-abelian affine Toda equation (\ref{5.14}) takes the  form,
\begin{eqnarray}
\pa_- ( g_1^{-1}\pa_+ g_1)&=&
\beta^t g_2^{-1} \bar\beta g_1 -g_1^{-1} \bar\alpha^t (g_{2}^t)^{-1}\alpha,
\nonumber\\
\pa_-( g_2^{-1}\pa_+ g_2) &=& D_0 g_3^{-1} \bar D_0 g_2
-g_2^{-1} \bar\beta g_1 \beta^t -g_2^{-1} \bar\alpha g_1^{-1}\alpha^t,
\nonumber\\
\pa_-( g_k^{-1}\pa_+ g_k) &=& D_0 g_{k+1}^{-1} \bar D_0 g_k
-g_k^{-1} \bar D_0 g_{k-1} D_0 , \qquad  2<k\leq p+1,
\label{T10}\end{eqnarray}
where $g_{p+2}^{-1}=g_{p}^t$.
The  conformal Toda equation corresponding to equation (6.17)
can be obtained by dropping the terms  containing $\alpha$
and $\bar\alpha$.
The simplest version of equation (6.17)
 arises for the Lie algebra $D_4$, and  describes a $GL(2)$ valued
field $g_2$ interacting
with two ``scalars''   $g_1\in GL(1)$ and $g_3\in O(2)$.

\newpage

\section{Conclusion}

In this paper we studied  a class of
generalized KdV hierarchies associated by Drinfeld-Sokolov
reduction to  regular semisimple elements of grade one in the
non-twisted loop algebras.
We made use of the fact that the classification
of the graded regular semisimple elements in a loop algebra $\ell(\G)$
can be reduced to the known \cite{Sp} classification of
the regular conjugacy classes in the Weyl group ${\bf W}(\G)$
of the underlying simple Lie algebra $\G$.
The regular conjugacy classes in ${\bf W}(\G)$
parametrize the non-equivalent Heisenberg subalgebras of $\ell(\G)$
containing graded regular semisimple elements.
Restricting  our attention to the  {\em classical}
simple Lie algebras,  we exhibited a relationship between
the regular conjugacy classes
in ${\bf W}(\G)$ and certain
$sl_2$ subalgebras of $\G$.

Let $[w]\subset {\bf W}(\G)$ be a regular conjugacy class
of order $m$ for $\G$ a  classical simple Lie algebra.
We have seen that there exists a lift $\hat w$ of
a representative  $w\in [w]$ that takes  the form
$\hat w=\exp\left(2i \pi {\rm ad } I_0/m\right)$ in such a way
that $I_0$ is the semisimple element (``defining vector'')
of an $sl_2$ subalgebra of $\G$ for which the largest eigenvalue
of ${\rm ad }I_0$ is $(m-1)$.
Any regular element $\Lambda$ of minimal positive grade from
the corresponding Heisenberg subalgebra has the form
$\Lambda=(C_1 + \lambda C_{-(m-1)})$,
where $[I_0, C_k]=kC_k$ and $C_1$
can be included in  an $sl_2$ subalgebra containing $I_0$.
The grade of $\Lambda$ is one with respect to the grading
operator $d_{m, I_0}=m \lambda {d\over d\lambda} + {\rm ad} I_0$.

In the appendix it will be observed that the same
relationship  is valid
between arbitrary {\em regular primitive} conjugacy classes
in the  Weyl group and certain $sl_2$ embeddings for {\em arbitrary}
simple Lie algebras.
For a non-primitive regular conjugacy class $[w]$ in the Weyl group of an
exceptional simple Lie algebra different from $G_2$,
in some cases the order of $w\in [w]$ is smaller than the
largest spin plus one  with respect to the $sl_2$
associated to $[w]$.

Applying the above group theoretic  results,  we provided a link between
the generalized KdV  hierarchies and $\W$-algebras and
made a step towards obtaining
a more concrete description of the KdV systems.
In particular, we derived  Gelfand-Dickey type Lax operators for
the KdV systems associated to  grade one regular elements from
such Heisenberg subalgebras that are
contained in a regular reductive subalgebra of a classical
Lie algebra $\G$ comprising $A$ and $C$ type simple factors.
In these  cases the generalized KdV systems turned out to be discrete
reductions of systems related  to $gl_n$
having Lax operators of the form given in (1.3) and (1.4).

The most interesting non-principal case occurring for the classical
Lie algebras appears to be given by the
regular primitive conjugacy class
$(\bar p,\bar p)\subset {\bf W}(D_{2p})$,
since the corresponding Heisenberg subalgebra is not contained in a
regular reductive subalgebra.
It is an intriguing question  whether the generalized  KdV
system  associated to a  grade one regular element  with the aid of
Drinfeld-Sokolov reduction admits a Gelfand-Dickey type pseudo-differential
operator model in this case or not.
Such a model is usually not hard to
find using the elimination procedure,
but for $(\bar p,\bar p)\subset {\bf W}(D_{2p})$ we did not succeed until now.
The corresponding non-abelian affine Toda system presented
in Section 6 would also deserve further investigation.

In this study we used the interplay between the homogeneous grading
and the grading given by $d_{m,I_0}$  to define the constraints
on the first order  differential
operator ${\cal L}=\pa +j + \Lambda$ containing  the dynamical variables.
It is known  \cite{DS,Wi, McI, Prin1} that there
are more general possibilities: {\em i)}
the $d_{m,I_0}$  grading can be  replaced by an arbitrary  grading in which
$\Lambda$ has definite grade; {\em ii)}  the
homogeneous grading can be replaced  by another standard
grading (associated to an appropriate vertex of the extended Dynkin
diagram) or a grading
interpolating between a standard grading and the
grading in which $\Lambda$ has definite grade.
See also the remark at the end of Section 2.
It would be interesting  to further explore these more general
possibilities  for obtaining  KdV and partially modified KdV systems,
which are related to the same basic set of modified KdV systems by
different Miura maps \cite{DS,Wi, McI, Prin1}.

We wish to remark  that in some cases the partially
modified systems correspond to partial factorizations of
a Lax operator that
can be factorized into factors of order one, not unlike the
example when say a fourth order KdV operator $L$
is partially factorized into operators of order two
according to
$L=(\pa +\theta_1)(\pa +\theta_2)(\pa + \theta_3)(\pa+\theta_4)=L_1L_2$ with
$L_1=(\pa +\theta_1)(\pa + \theta_2)$ and
$L_2=(\pa +\theta_3)(\pa +\theta_4)$.

We have  restricted our attention to regular elements of {\em minimal} grade.
According to an argument in \cite{Prin1,Prin2}, the systems associated to
regular elements of higher grade in a certain sense should not be  new,
although the Hamiltonian aspect of this claim is not well understood.

Perhaps the most serious  limitation of the present work is that we excluded
``type II'' systems,
that is systems associated to graded {\em non-regular}  semisimple
elements of $\ell(\G)$,
from the outset.
It is an important open problem  to classify the  gradings that admit graded
semisimple elements for which Drinfeld-Sokolov reduction is possible
in the sense that polynomial ``DS gauges'' exist.
Some results on type II systems including interesting
examples can be found in  \cite{FoKu,McI+,McI, SdeC,FM}.
In particular, it was recently shown in \cite{SdeC} that the phase space
of the partially modified systems
contains standard $\W$-algebras coupled together by the dynamics
in both type I and type II cases
subject to a certain non-degeneracy condition.

It is worth noting that the regular conjugacy classes
in the groups obtained as extensions
of the Weyl groups by diagram automorphisms
have been also classified in \cite{Sp},
which is relevant for constructing  generalized KdV
and affine Toda systems based on the twisted loop algebras.

To conclude, we think the general framework of the Drinfeld-Sokolov
approach is now reasonably clear but further work would be needed
to fully classify  the integrable  hierarchies that can be obtained from
this approach.
For instance, it would be of some interest to further explore the
KdV systems that may be defined using arbitrary  grade one regular
semisimple elements and arbitrary  standard  gradings
and  type II systems
would also deserve closer attention.

\medskip

\bigskip

\bigskip

\bigskip

\noindent
{\large \bf  Acknowledgements.}
We wish to thank D.~Olive, P.~Sorba, F.~Toppan and J.~Underwood
for discussions.
We are grateful to I.R.~McIntosh for making available ref.~\cite{McI}
and to a referee for drawing our attention to ref.~\cite{Sp}.
We are indebted to A.~Honecker, I.~Marshall and I.~Tsutsui
for  comments and suggestions concerning  the manuscript.

\newpage

\appendix

\setcounter{equation}{0}

\renewcommand{\theequation}{A.\arabic{equation}}

\section{Canonical $sl_2$ for any regular primitive conjugacy class}

The purpose of this appendix is to present a property of the
regular primitive conjugacy classes in ${\bf W}(\G)$ that generalizes
the celebrated relationship \cite{Kost} between the Coxeter class
and the principal $sl_2$ subalgebra of $\G$.
We find this  relationship by collecting  known results in the
literature.
A larger set  of regular conjugacy classes
enjoying  the attractive features of this relationship
(properties 1--7 below) will be also pointed out.

Let $\G$ be an arbitrary simple Lie algebra.
The primitive (semi-Coxeter) conjugacy classes in ${\bf W}(\G)$
are the building blocks of the general conjugacy classes \cite{cart}
and the {\em regular primitive} conjugacy classes are the
building blocks of the general regular conjugacy classes.
The Coxeter class, whose Carter diagram \cite{cart} is the Dynkin
diagram of $\G$, is the only primitive conjugacy class
for the algebras of $A$, $B$, $C$ and $G_2$ type.
The other primitive conjugacy classes  can be uniquely
labelled by the Carter diagrams $D_l(a_i)$ for $i=1,\ldots, [l/2]-1$,
$F_4(a_1)$, $E_6(a_i)$ for $i=1,2$, $E_7(a_i)$ for $i=1,\ldots, 4$
and $E_8(a_i)$ for $i=1,\ldots, 8$.
The Coxeter class is always regular.
Comparing the characteristic polynomials of the primitive
conjugacy classes given in \cite{cart} with those of the regular
conjugacy classes given in \cite{Sp}, it can be seen that the
other {\em regular primitive} conjugacy classes are
$D_{2k}(a_{k-1})\sim (\bar k,\bar k)$ in ${\bf W}(D_{2k})$ for
$k=2,3,\ldots$, and
\begin{equation}
F_4(a_1),\quad
E_6(a_1),\,\, E_6(a_2),
\quad
E_7(a_1),\,\, E_7(a_4),
\quad
E_8(a_i)\quad \hbox{for}\quad i=1,2,3,5,6,8.
\label{A.1}\end{equation}
Putting together results of \cite{Kost,Sp,leur,bouw},
we notice the validity of the following statement.

\medskip
\noindent
{\bf Theorem A.}  {\em Let $[w]\subset {\bf W}(\G)$ be an
arbitrary regular primitive conjugacy class of order $N$.
Then there exists a lift $\hat w$  of $w\in [w]$ given by
an inner automorphism of $\G$  that has the form
\begin{equation}
\hat w=\exp\left(2i\pi {\rm ad}I_0/N\right),
\label{A.2}\end{equation}
where $I_0$ is the semisimple
element
of an $sl_2$ subalgebra of $\G$, $[I_0, I_\pm]=\pm I_\pm$, $[I_+, I_-]=2I_0$,
such that

\noindent\quad 1.
The largest eigenvalue of ${\rm ad} I_0$ equals $(N-1)$.

\noindent \quad 2.
There are no singlets in the $sl_2$ decomposition of $\G$.

\noindent \quad 3.
Only integral eigenvalues of ${\rm ad} I_0$ occur.
}
\medskip

\noindent
{\em Verification.} The case of the Coxeter class  is
due to  Kostant \cite{Kost}.
The characteristic  diagrams \cite{dyn} of the $sl_2$ embeddings
corresponding\footnote{In \cite{Sp} there is a
misprint in the  diagram of the $sl_2$ with  Dynkin index 280
that corresponds to $E_8(a_5)$.}
to the conjugacy classes
\begin{equation}
E_6(a_1),\quad
E_7(a_1),\quad
E_8(a_1), \,\, E_8(a_2),\,\, E_8(a_5)
\label{A.3}\end{equation}
are given in Table 11 of Ref.~\cite{Sp}, where the statement
is proved concerning  these cases.
(See also remarks iii) and  vii) below.)
In the  algebras of $E$ type, the
``shift vector'' $\gamma_s\in \G$ defining a so called canonical lift
 of a representative $w\in [w]$
was determined  by Bouwknegt \cite{bouw}
for  all  conjugacy classes $[w]\subset {\bf W}(\G)$.
For the definition  and for the  rather complex
method whereby $\gamma_s$ was obtained, see \cite{bouw}.
In the case of
the primitive conjugacy classes this canonical lift
takes the form $\hat w =\exp\left(2i\pi {\rm ad} \gamma_s/N\right)$.
Comparing the tables of \cite{bouw} with the tables of Dynkin \cite{dyn},
one can verify  that $\gamma_s\in \G$  coincides with the
defining vector of an $sl_2$ embedding {\em if  and only if} the
conjugacy class is regular.
The  $sl_2$ embeddings corresponding to the conjugacy classes
\begin{equation}
E_6(a_2),\quad E_7(a_4),\quad E_8(a_3),\,\,\, E_8(a_6),\,\, E_8(a_8)
\label{A.4}\end{equation}
are in this way identified as those with Dynkin index \cite{dyn}
\begin{equation}
36,\quad 39,\quad 184,\,\,\, 120,\,\, 40,
\label{A.5}\end{equation}
respectively.  Properties 1, 2, 3 can be checked.
In the $D_{2k}(a_{k-1})$ cases the lift satisfying the statement
of the theorem
was determined in \cite{leur}, as we have discussed in Subsection 4.2
using the alternative parametrization
$D_{2k}(a_{k-1})\sim (\bar k,\bar k)$.
The remaining $F_4(a_1)$ case results from the $E_6(a_2)$ case by
applying the canonical diagram automorphism $\tau$ of $E_6$, whose fixed
point set is $F_4$.
This is similar to an appropriate representative of the
Coxeter class of $E_6$ reducing to a representative of the Coxeter
class of $F_4$ on the fixed point set of $\tau$, which is well-known.
In fact, $E_6(a_2)$ and $F_4(a_1)$ can be represented by the squares
of the respective Coxeter elements. The $sl_2$ embedding associated
to $E_6(a_2)$ by the theorem  is the principal $sl_2$ in the regular
subalgebra
$(A_5+A_1)\subset E_6$, which is the same as the principal $sl_2$
in the regular subalgebra $(C_3+A_1)\subset F_4$.
Using also Lemma 9.5 of Springer \cite{Sp}, we can conclude that
the latter $sl_2$ subalgebra of $F_4$, having Dynkin index 36, satisfies
the statement of the theorem for $F_4(a_1)$. {\em Q.E.D.}

\medskip

Any representative of a regular primitive
conjugacy class  $[w]\subset {\bf W}(\G)$ of order $N$ has \cite{Sp} a regular
semisimple eigenvector associated to the  eigenvalue
$\omega_N:=\exp\left(2i\pi/N\right)$.
For the lift $\hat w$ given in the theorem, any semisimple
eigenvector $H$ of eigenvalue $\omega_N$ has the form
\begin{equation}
H=C_1 + C_{-(N-1)}
\quad\hbox{with}\quad
C_k\neq 0, \quad
[I_0, C_k]=k C_k.
\label{A.6}\end{equation}
We have the following consequence of the theorem.

\medskip
\noindent
{\bf Corollary A.} {\em Let $\hat w$ be the lift of a regular primitive
conjugacy class $[w]\subset {\bf W}(\G)$ given in the theorem
and $H$ in (A.6) be a  regular
semisimple eigenvector of $\hat w$  with eigenvalue $\omega_N$.
Let ${\cal H}_H\subset \G$ be the Cartan subalgebra defined
as the centralizer of $H$. Then

\noindent \quad 4.
The restriction of $\hat w$ to ${\cal H}_H$ acts as a representative
of the  conjugacy class $[w]\subset {\bf W}(\G)$.

\noindent \quad 5.
$I_0$ and $C_1$ can be completed to an $sl_2$ subalgebra of $\G$.
}
\medskip

\noindent
{\em Proof.} Since $\hat w$ maps ${\cal H}_H$ to itself it defines
a representative of a conjugacy class in ${\bf W}(\G)$.
This conjugacy class is obviously  regular
and has order $N$. Property 4 follows since there can be only one
regular conjugacy classes of a given order \cite{Sp}.
To show property 5, notice that
${\rm dim\,} \G_{-1}^{I_0}= {\rm dim\,} \G_{0}^{I_0}$ by property 2 in
the theorem. Further notice  that
\begin{equation}
{\rm Ker}\left( {\rm ad} C_1\right) \cap \G^{I_0}_{<0}=\{ 0\}
\label{A.7}\end{equation}
by property 1 and by the assumption that $H$ in (A.6)  is
regular semisimple. {\em Q.E.D.}

\medskip
Let $I_0$, $I_\pm$ be the $sl_2$ subalgebra given in the theorem
and $C_{-(N-1)}$ some element of $\G^{I_0}_{-(N-1)}$.
Note that for $I_0$ given  $I_\pm$ are  not unique.
Springer \cite{Sp} has also shown  the following:
\medskip

{\em \noindent \quad
6. If $(I_+ + C_{-(N-1)})$ is semisimple then it is regular semisimple.

\noindent \quad
7. There exists $C_{-(N-1)}$ such that $(I_+ + C_{-(N-1)})$ is
regular semisimple.
}

\medskip
We wish to make some further remarks on the
``canonical correspondence''  between $sl_2$ embeddings and
regular primitive conjugacy classes established above.

\smallskip
\noindent
i) The shift vector  defining the canonical lift \cite{bouw,leur}  of a
primitive conjugacy class in ${\bf W}(\G)$
determines an $sl_2$ embedding {\em only} if
the conjugacy class is regular.

\smallskip
\noindent
ii) The $sl_2$ corresponding to a regular
primitive (semi-Coxeter) conjugacy class is {\em not} always a singular
(semi-principal)  $sl_2$.

\smallskip
\noindent
iii) The principal $sl_2$ and the $sl_2$ subalgebras
corresponding to the conjugacy classes
in (A.3) satisfy \cite{Sp}  in addition to properties 2, 3  also
the property that there occurs only one  triplet in the $sl_2$
decomposition of $\G$.
There exists only one additional $sl_2$ embedding with these properties,
corresponding to the regular embedding $B_4\subset F_4$.
The multiplicity of the largest  spin $sl_2$ multiplet in $\G$ is also one
in these cases.

\smallskip
\noindent
iv) Relation (A.2) alone would {\em not} determine uniquely the
conjugacy class of the $sl_2$ generator $I_0$ (think of non-conjugate
powers of a Coxeter element).
It may be checked that (A.2) together with property 1 does so.

\smallskip
\noindent
v) The shift vector determined in \cite{bouw} for all the conjugacy
classes in
${\bf W}(E_{6,7,8})$  associates  an $sl_2$ embedding
to every regular
conjugacy class. Property 1 is {\em not} always satisfied, but the
weaker requirement $\G_1^{\hat w}=\G_0^{I_0}$ holds for the eigenspaces
of $\hat w$ and ${\rm ad}I_0$ with the respective eigenvalues.
If property 1 is not satisfied then
it might be necessary to modify
the definition of the Drinfeld-Sokolov reduction used in Section 5,
because relation (5.2), $[C_-, {\cal G}^{I_0}_{<0}]=\{0\}$, is then
{\em not} guaranteed to hold for the grade one regular semisimple
element $\Lambda=(I_+ + \lambda C_-)$.

\smallskip
\noindent
vi) There exist a few other  $sl_2$ embeddings and non-primitive
regular conjugacy classes in the Weyl group of a simple Lie algebra $\G$
for which {\em all} of the above presented equations and properties 1--7
hold true as well.
These conjugacy classes in ${\bf W}(\G)$
 are given  by the following Carter diagrams:
\begin{equation}
D_{2k}(a_{k-1})\in {\bf W}(B_{2k})\quad\hbox{for}\quad k\geq 1,\quad
A_2\in {\bf W}(G_2),\quad
B_4\in {\bf W}(F_4), \quad
D_4(a_1)\in {\bf W}(F_4),
\label{A.8}\end{equation}
where for $k=1$ we use the definition
 $D_{2}(a_0):=(\bar 1,\bar 1)$.
The corresponding $sl_2$ is obtained by taking
the  semi-principal or principal $sl_2$
embedding in the respective regular simple subalgebras of maximal rank,
\begin{equation}
D_{2k}(a_{k-1})\subset B_{2k} \quad\hbox{for}\quad k\geq 1,\qquad
A_2\subset G_2,\qquad
B_4\subset F_4, \qquad
D_4(a_1)\subset F_4,
\label{A.9}\end{equation}
where $D_{2k}(a_{k-1})$ denotes the
semi-principal $sl_2$ subalgebra in $D_{2k}$ described
in Subsection 4.2.
For the alert reader, we note that
$D_4(a_1)\subset F_4$ is the $sl_2$ of Dynkin index 12,
although this labelling of it is missing in the table of \cite{dyn}.

\smallskip
\noindent
vii) Springer \cite{Sp} studied the  correspondence between
$sl_2$ embeddings and regular conjugacy classes
in the Weyl group using
in addition to  (A.2) and properties 1, 2, 3 the assumption
that there exists a regular semisimple eigenvector of $\hat w$
given by (A.2) of the form in (A.6).
It can be checked that  the $sl_2$ subalgebras
corresponding to the regular primitive conjugacy
classes together with those in (A.8) yield  the  {\em exhaustive set}
for which these assumptions are satisfied.
In \cite{Sp} the strong additional assumption that
the decomposition of $\G$ under the $sl_2$ contains
only one triplet was used to ensure the existence
of a regular semisimple eigenvector.

\medskip
In the above  we have described a canonical correspondence
between the regular primitive conjugacy classes in the Weyl group
and certain associated $sl_2$ embeddings.
The correspondence enjoys a set of attractive properties,
which are shared by certain other regular non-primitive conjugacy classes,
given in (A.8), and corresponding $sl_2$ embeddings.
Some further nice properties valid in these cases can be
found in \cite{Sp}.
This correspondence enhances our understanding of the
classification of integrable hierarchies associated to regular
conjugacy classes in the Weyl group
and could be further exploited in more detailed studies of these
systems.

\newpage



\begin{thebibliography}{34}

\bibitem[1]{DS}
V.G.\ Drinfeld and  V.V.\ Sokolov,
Sov.\ Math.\ Dokl.\ {\bf 23} (1981) 457;
J. Sov. Math. {\bf 30} (1985) 1975.


\bibitem[2]{GD}
I.M.\ Gelfand and L.A.\ Dickey,
Funct.\ Anal.\ Appl.\ {\bf 10:4} (1976) 13;
Funct.\ Anal.\ Appl.\ {\bf 11:2} (1977) 93.

\bibitem[3]{Di}
L.A.\ Dickey,
{\em Soliton Equations and Hamiltonian Systems},
Adv.\ Ser.\ Math.\ Phys., Vol.\ 12,
World Scientific, Singapore,  1991.


\bibitem[4]{Ad}
M.\ Adler,
Invent.\ Math.\ {\bf 50:3} (1979) 219.


\bibitem[5]{RSTS}
A.G.\ Reyman and M.A.\ Semenov-Tian-Shansky,
Funct.\  Anal.\  Appl.\  {\bf 14} (1980) 146.

\bibitem[6]{kac}
V.G.\ Kac, {\sl Infinite Dimensional Lie Algebras},
Cambridge University Press, Cambridge, 1985.


\bibitem[7]{Wi}
G.W.\ Wilson,
Ergod.\ Th.\ and Dynam.\ Sys.\ {\bf 1} (1981) 361.

\bibitem[8]{LS}
A.N.\ Leznov and M.V.\ Saveliev,
Commun.\ Math.\ Phys.\ {\bf 89} (1983) 59.

\bibitem[9]{Oli}
D.\ Olive and N.\ Turok,
Nucl.\ Phys.\ {\bf B257} (1985) 277.

\bibitem[10]{under}
J.W.R.\ Underwood,
London preprint
IC/TP/92-93/30,  hep-th/9304156;
{\em Toda Theories as  Model Integrable Systems}, PhD Thesis,
Dept.\ Phys.,
Imperial College, London, 1993.



\bibitem[11]{McI}
I.R.\ McIntosh,
{\em An Algebraic Study of Zero Curvature Equations},
PhD Thesis, Dept.\ Math., Imperial College, London, 1988.


\bibitem[12]{Prin1}
M.F.\ de Groot, T.J.\ Hollowood and J.L.\ Miramontes,
Commun.\ Math.\ Phys.\ {\bf 145} (1992) 57.


\bibitem[13]{Prin2}
N.J.\ Burroughs, M.F.\ de Groot, T.J.\ Hollowood and J.L.\ Miramontes,
Commun.\ Math.\ Phys.\ {\bf 153} (1993) 187;
Phys.\ Lett.\ {\bf B277} (1992) 89.


\bibitem[14]{Prin3}
T.J.\ Hollowood and J.L.\ Miramontes,
Commun.\ Math.\ Phys.\ {\bf 157} (1993) 99.

\bibitem[15]{SdeC}
C.R.\ Fern\'andez-Pousa, M.V.\ Gallas, J.L.\ Miramontes and J.S.\ Guill\'en,
Santiago de Compostela preprint, US-FT/13-94, hep-th/9409016.


\bibitem[16]{KP}
V.G.\ Kac and D.H.\ Peterson,
pp.~276-298 in:~Proc.\ of Symposium on Anomalies, Geometry and Topology,
W.A.\ Bardeen and A.R.\ White (eds.),
World Scientific, Singapore, 1985.


\bibitem[17]{FHM}
L.\ Feh\'er, J.\ Harnad and I.\ Marshall,
Commun.\ Math.\ Phys.\ {\bf 154} (1993) 181.

\bibitem[18]{FM}
L.\ Feh\'er and  I.\ Marshall,
Swansea preprint, SWAT-95-61, hep-th/9503217.

\bibitem[19]{ten}
F.\ ten Kroode and  J.\ van de Leur,
Commun.\ Math.\ Phys.\ {\bf 137} (1991) 67.

\bibitem[20]{Cheng}
Yi Cheng,
J.\ Math.\ Phys.\ {\bf 33} (1992) 3774.

\bibitem[21]{OS}
W.\ Oevel and  W.\ Strampp,
Commun.\ Math.\ Phys.\ {\bf 157} (1993) 51.

\bibitem[22]{Deck}
A.\ Deckmyn,
Phys.\ Lett.\ {\bf B298} (1993) 318.


\bibitem[23]{Dic}
L.A.~Dickey,
Oklahoma preprint, hep-th/9407038.

\bibitem[24]{Bon}
L.\ Bonora, Q.P.\ Liu and C.S.\ Xiong,
Bonn preprint, BONN-TH-94-17, hep-th/9408035.

\bibitem[25]{Ara}
H.\ Aratyn, J.F.\ Gomes and  A.H.\ Zimerman,
Chicago preprint, UICHEP-TH/93-10, hep-th/9408104.


\bibitem[26]{cart}
R.\W.\ Carter,
Comp.\ Math.\ {\bf 25} (1972) 1.


\bibitem[27]{Sp}
T.A.\ Springer,
Inventiones math.\ {\bf 25} (1974) 159.


\bibitem[28]{Zam}
A.B.\ Zamolodchikov,
Theor.\ Math.\ Phys.\ {\bf 65} (1985) 1205.

\bibitem[29]{Doug}
M.\ Douglas,
Phys.\ Lett.\ {\bf B238} (1990) 176.

\bibitem[30]{bais}
F.A.\ Bais,  T.\ Tjin and P.\ van Driel,
Nucl.\ Phys.\ {\bf B357} (1991) 632.

\bibitem[31]{rep}
L.\ Feh\'er, L.\ O'Raifeartaigh, P.\ Ruelle, I.\ Tsutsui and A.\ Wipf,
Phys.\ Rep.\  {\bf 222} (1992) 1.


\bibitem[32]{BS}
P.\ Bouwknegt and  K.\ Schoutens,
Phys.\ Rep.\ {\bf 223} (1993) 183.


\bibitem[33]{Mor}
A.\ Morozov,
Amsterdam preprint, IFTA-93-10 (1993).


\bibitem[34]{dyn}
E.B.\ Dynkin,
Amer.\ Math.\ Soc.\ Transl.\ {\bf 6 [2]} (1957) 111.


\bibitem[35]{Kost}
B.\ Kostant,
Am.\ J.\ Math.\ {\bf 81} (1959) 973.

\bibitem[36]{segal}
A.\ Pressly and G.\ Segal, {\sl Loop Groups},
Oxford University Press, 1986.

\bibitem[37]{fons}
F.\ ten Kroode,
{\em Affine Lie Algebras and Integrable Systems},
PhD Thesis, University of Amsterdam, 1988.


\bibitem[38]{hel}
S.\ Helgason, {\sl Differential Geometry, Lie Groups and Symmetric
Spaces},  Academic Press,  New York, 1978.



\bibitem[39]{leur}
F.\ ten Kroode and J.\  van de Leur,
Comm.~in Algebra {\bf 20} (1992) 3119.

\bibitem[40]{cart2}
R.W.\ Carter and  G.B.\ Elkington,
Journal of Algebra {\bf 20} (1972) 350.


\bibitem[41]{bouw}
P.\ Bouwknegt,
J.\ Math.\ Phys.\ {\bf 30} (1989) 571-584.




\bibitem[42]{jac}
N.\ Jacobson,
{\sl Lie Algebras},
Wiley-Interscience, New York, 1962.




\bibitem[43]{RSTS2}
A.G.\ Reyman and M.A.\ Semenov-Tian-Shansky,
Phys.\ Lett.\ {\bf A130} (1988) 456.

\bibitem[44]{bal}
J.\ Balog, L.\ Feh\'er, L.\ O'Raifeartaigh, P.\ Forg\'acs and A.\ Wipf,
Ann.\ Phys.\ (N.Y.) {\bf 203} (1990) 76.



\bibitem[45]{FoKu}
A.P.\ Fordy and  P.P.\ Kulish,
Commun.\ Math.\ Phys.\ {\bf 89} (1983) 427.


\bibitem[46]{McI+}
I.R.\ McIntosh,
J.\ Math.\ Phys.\ {\bf 34} (1993) 5159.



\end{thebibliography}
\end{document}